\documentclass{aastex63}

\usepackage{amsmath,bm}
\usepackage{rotating}
\usepackage[caption=false]{subfig}

\newcommand{\swift}{\textit{Swift}}
\newcommand{\fermi}{\textit{Fermi}}

\shorttitle{NITRATES}
\shortauthors{DeLaunay \& Tohuvavohu}
\graphicspath{{./}{figures/}}

\begin{document}

\title{Harvesting BAT-GUANO with NITRATES (Non-Imaging Transient Reconstruction And TEmporal Search): Detecting and localizing the faintest GRBs with a likelihood framework
}

\author[0000-0001-5229-1995]{James DeLaunay}
\affiliation{Department of Physics, Pennsylvania State University, University Park, PA 16802, USA}
\affiliation{Center for Multimessenger Astrophysics, Institute for Gravitation and the Cosmos, Pennsylvania State University, University Park, PA 16802, USA}
\affiliation{Department of Physics \& Astronomy, University of Alabama, Tuscaloosa, AL 35487, USA}

\author[0000-0002-2810-8764]{Aaron Tohuvavohu}
\affiliation{David A. Dunlap Department of Astronomy \& Astrophysics, University of Toronto, Toronto, ON, Canada}



\begin{abstract}
The detection of the gravitational wave counterpart GRB 170817A, underluminous compared to the cosmological GRB population by a factor of 10,000, motivates significant effort in detecting and localizing a dim, nearby, and slightly off-axis population of short GRBs. Swift/BAT is one of the most sensitive GRB detector in operation, and the only one that regularly localizes GRBs to arcminute precision, critical to rapid followup studies. However, the utility of BAT in targeted sub-threshold searches had been historically curtailed by the unavailability of the necessary raw data for analysis. The new availability of time-tagged event (TTE) data from the GUANO system \citep{GUANO}, motivates renewed focus on developing sensitive targeted search analysis techniques to maximally exploit these data. While computationally cheap, we show that the typical coded-mask deconvolution imaging is limited in its sensitivity due to several factors. We formalize a maximum likelihood framework for the analysis of BAT data wherein signals are forward modelled through the full instrument response, and -- coupled with the development of new response models -- demonstrate its superior sensitivity to typical imaging via archival comparisons, injection campaigns, and, after implementing as a targeted search, a large number of low-latency GRB discoveries and confirmed arcminute localizations to date. We also demonstrate independent localization of some out-of-FOV GRBs for the first time. NITRATES's increased sensitivity boosts the discovery rate of GRB 170817A-like events in BAT by a factor of at least $3-4$x, along with enabling joint analyses and searches with other GRB, GW, neutrino, and FRB instruments. We provide public access to the response functions and search pipeline code.
\end{abstract}



\section{Introduction} \label{sec:intro}
The prompt arcminute localization of prompt Gamma-Ray Burst (GRB) emission has been crucial to finding their afterglows, redshifts, and making inferences on progenitor populations, along with all other science that requires follow-up \citep{GRBreviewberger,GRBreviewlevan}.  To date, this has only been regularly accomplished via the use of coded aperture mask instruments, which `image' the hard X-ray and gamma-ray sky via a `mask' whose shadow, cast onto the detector plane, spatially encodes the incident radiation in a unique way for each direction in the coded field-of-view (FOV). The Burst Alert Telescope (BAT; \citealt{BAT}) onboard the Neil Gehrels \swift\ Observatory (hereafter: \swift; \citealt{SwiftMain}) is the largest such coded aperture imager yet launched, with a $\sim 2.2$ steradian coded FOV, an on-axis effective area of $2700$ cm$^2$ at launch, and the capability to localize sources to $<3$ arcminute precision. This combination of high effective area,  large FOV, and precise localization has enabled BAT to discover and localize $>1500$ GRBs to date, including the detection of the most distant \citep{9.4grb}, and longest \citep{grb111209a} GRBs, as well as enabling the rapid followup that has allowed the discovery of $>1400$ GRB afterglows, and several kilonova candidates, with detections spanning the entire electromagnetic spectrum.

The discovery of gravitational waves (GW) from a neutron star merger \citep{GW170817} associated with an extremely under-luminous short GRB (GRB 170817A; \citealt{GWGRB}, \citealt{GBMgrb170817A}, \citealt{INTEGRALgrb170817}) has motivated significant effort to find faint short GRBs, and in particular the further development of more sensitive, targeted, GRB searches around the times of GW events \citep{GBMtarg19}. For example, the \fermi\ Gamma-ray Burst Monitor (GBM) offline untargeted search effectively doubles their short GRB detection rate compared to onboard triggers, and the targeted search similarly increases their range for GRB 170817A-like events by $\sim 50\%$ \citep{GBMsubthresh}. These sensitive searches have crucially depended on the availability of continuous time and energy tagged photon count data (Time-Tagged Event data, hereafter: TTE or event data) from \fermi/GBM.

However, similar gains have not been demonstrated for \swift/BAT, despite its intrinsically superior sensitivity and localization capability. Throughout the vast majority of the \swift\ mission, the discovery and localization of GRBs has typically been limited to those found by the real-time analyses \citep{triggeralg} running onboard the spacecraft. This has been a result of the absence of TTE data on the ground for analysis, due to the combination of BAT's high effective area, and the insufficiency of the \swift\ downlink bandwidth to carry the large volume of event data to the ground. With a few notable and rare exceptions (eg \citealt{BATSS}), this resulted in dramatically limited yield from BAT ground analyses, and a frustrating inability to exploit the full capabilities of the instrument.

However, the recent development of the Gamma-ray Urgent Archiver for Novel Opportunities (GUANO; \citealt{GUANO}) has effectively solved the problem of the absence of data for targeted searches, now successfully retrieving $\sim 200$ second windows of BAT event data around the times of compelling astrophysical events multiple times per day with $>90\%$ success rate, as long as it is triggered by an external instrument within a few minutes of the time of interest. The availability of the event data of interest on the ground motivates the development of novel analysis techniques, in order to fully exploit the unique sensitivity and localization capabilities of the BAT instrument.

Here we introduce the Non-Imaging Transient Reconstruction and TEmporal Search (NITRATES), a new analysis technique for \swift/BAT event data, which allows the discovery and localization of significantly weaker GRBs in the BAT data, with wide ranging implications for the GRB arcminute localization, and therefore followup, rate as well as multi-messenger searches. We begin in Section \ref{sec:BAT} by describing the BAT instrument and the available data. In Section \ref{sec:imaging} we review the conventional coded aperture imaging technique for GRB localization, and said technique's limitations. In Section \ref{LikeSect} we describe and formalize a likelihood-based forward modelling analysis framework. In Section \ref{sec:respmodels} we construct the full instrumental response models required for the forward folding. In Section \ref{sec:GRBsearch} we utilize the likelihood analysis in a targeted GRB search and construct a test statistic. In Section \ref{sec:SearchPipe} we describe the full implementation of the automated targeted search pipeline. In Section \ref{sec:sensitivity} we estimate the sensitivity of this search with signal injections of GRB 170817A and assess the increased recovery range. In Section \ref{SectDiscoveries} we give examples of GRB location and afterglow discoveries that were uniquely enabled by the sensitivity of this new analysis using data from GUANO. In Section \ref{sec:ImproveSearch} we discuss various prospects for further enhancements to the NITRATES analysis.
As a companion to this paper, we release the full end-to-end NITRATES pipeline code,\footnote{ \href{https://github.com/Swift-BAT/NITRATES}{https://github.com/Swift-BAT/NITRATES}} developed in Python, along with the full BAT instrumental response functions developed for this purpose.\footnote{\href{https://zenodo.org/communities/swift-bat}{https://zenodo.org/communities/swift-bat}}

\section{The Burst Alert Telescope}
\label{sec:BAT}
The difficulty of focusing photons with energies above $\sim15$ keV coupled with the requirement for survey telescopes to have a large instantaneous field-of-view (FOV) has driven the adoption and development of coded aperture mask instruments. The \swift/BAT is the largest such yet launched. The detector plane is comprised of 32,768 4 x 4 x 2 mm$^3$ CdZnTe photon counting detectors, making a $5200$ cm$^2$ detection area. Above this is a lead and composite mask made of 54,000 5 x 5 x 1 mm$^3$ lead tiles (in a 50\% open-closed fully random pattern) with total area of $2.7$ m$^2$. The mask sits one meter above the detector plane, yielding a $\sim2.2$ steradian FOV out to $10\%$ partial coding. A graded shield surrounds the detector plane and the mask, dramatically reducing the background, and ensuring that most photons incident on the detector must come through the mask holes (though at photon energies $>200$ keV even this shielding is transmissive). 

The photon counting detectors have $100\,\mu s$ relative timing accuracy, and tag detected counts with their respective detector ID, timestamps, and an energy in one of 4096 channels ranging from 15-500 keV (although the response has historically only been well calibrated up to $\sim$350 keV), with an energy resolution of $\sim5$ keV at 60 keV; this is the event data. The size of the mask cells and their distance from the detector plane sets an imaging PSF size of 17 arcminutes (FWHM), and a source position centroid accuracy of 1-3 arcminutes.

\begin{figure}
\centering
\begin{subfloat}
  \centering
  \includegraphics[width=.4\linewidth]{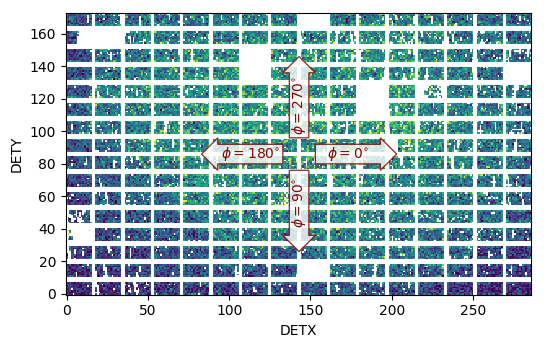}
\end{subfloat}
\begin{subfloat}
  \centering
  \includegraphics[width=.5\linewidth]{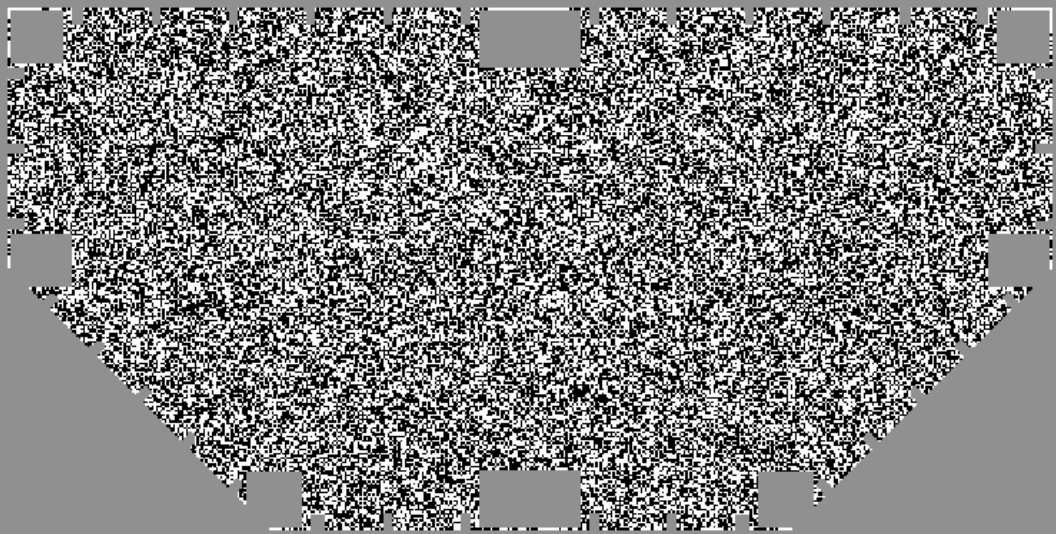}
\end{subfloat}
\linebreak
\begin{subfloat}
  \centering
  \includegraphics[width=.6\linewidth]{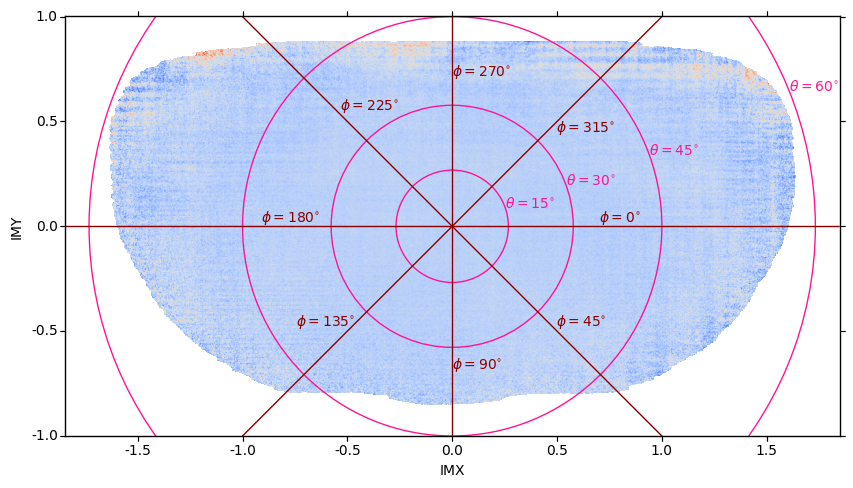}
\end{subfloat}
\caption{The top left plot is a detector plane image (DPI), showing the accumulated counts in each CZT detector across the full detector array. The DETX and DETY coordinates are shown on the axes and the directions of $\phi$ are shown via arrows. The $\theta=0$ direction points straight out of the page and through the mask. The top right plot shows the full mask image, black (white) indicates an closed (open) lead (composite) cell. The relative sizes of the DPI and mask are not to scale. Grey shows the uncoded regions, including structural supports for the mask. The bottom plot is a BAT sky image, showing the reconstructed mask-weighted counts for the entire, $\sim2.2$ steradian, coded field of view. The tangential plane coordinates, IMX and IMY are shown on the axes, and the spherical coordinates, $\theta$ and $\phi$ are plotted over the image.  }
\label{fig:Coords}
\end{figure}

The detector array lies in the X-Y plane of the instrument's coordinate system. Each detector has its own position in the array labeled with DETX and DETY. Spherical coordinates are used to define directions to a point in the sky relative to the instrument, where the zenith angle, $\theta$ is the angle from the direction normal to the detector plane, looking up through the mask, and the azimuthal angle, $\phi$ is $0^{\circ}$ along the x-axis and increases in the $-$y direction so that it is $90^{\circ}$ along the $-$y-axis. An example detector plane image is shown on the left in Figure \ref{fig:Coords}, where the DETX and DETY coordinates can be seen along with the $\phi$ directions. The direction of $\theta = 0^{\circ}$ is through the mask above the detector array and coming straight out of the page. For coded aperture imaging, it is convenient to work in tangential plane coordinates where the point spread function is constant in size. The coordinates commonly used in BAT sky images are IMX and IMY, where IMX = $\tan(\theta)\cos(\phi)$ and IMY = $\tan(\theta)\sin(-\phi)$. An example BAT sky image is shown on the bottom in Figure \ref{fig:Coords}, where the axes show the IMX and IMY coordinates and the $\theta$ and $\phi$ coordinates are plotted over the image. IMX and IMY are convenient to use inside the coded field of view, but are not really used outside of it, as they tend towards infinity as $\theta$ approaches $90^{\circ}$. These coordinate systems will be used throughout this paper. 

The Z-shield surrounding the sides and bottom of the instrument becomes highly transparent to photons with energies $>200$ keV, making the instrument sensitive to bright and hard GRBs from the entire unocculted sky, with hundreds of square centimeters of effective area even far outside its field of view, but without any ability to localize these sources via imaging. The Z-shield's graded thickness, along with the complexity of the surrounding spacecraft structure and X-ray and UV/Optical Telescope tubes leads to an anisotropic response for lines of sight outside of the coded area. The out of field of view response has not been well calibrated to date, meaning that, aside from participation in the Inter-Planetary Network (IPN) via timing, the BAT data has typically not been used for analysis of bursts detected from outside of the field of view -- with a few notable exceptions \citep{Palmer05, NatureMGF}.

\section{Coded aperture imaging}
\label{sec:imaging}
The event data: a list of counts tagged with detector ID, energy, and time can be manipulated and processed to produce a variety of data products, including sky images, mask-weighted (background subtracted) light curves of single sources, and spectra. Here we review the production of sky images, as this is the canonical approach for finding and localizing GRBs with coded aperture instruments.

The multi-dimensional (time, detector location, energy) event data can be arbitrarily binned and flattened into a 2D detector plane image (DPI, eg Figure \ref{fig:Coords}
 left panel), which is a histogram of counts per detector mapped onto the detector's location in the detection plane. This is accomplished by choosing a time range of interest, typically while the GRB is active, (and over which the attitude of the instrument is stable), and an energy range of interest, and binning along these dimensions. For GRB searches the time range is typically chosen by searches running in rates space, which attempt to identify the time window that will maximize the image SNR \citep{triggeralg}.

The counts per detector $i$ in this DPI (shadowgram) produced by a point source of mono-energetic flux $S$ can be calculated as
\begin{equation}
    R_i=S\cdot f_{\rm trans} \cdot \{f_i+(1-f_i)\cdot f_{Pb}\}\cdot A_{\rm eff} + B 
    \label{eq:rawcounts}
\end{equation}
where $f_{Pb}$ is the transmission of the lead mask tiles and $f_{\rm trans}$ the transmission of any other passive materials between the source and the detector. $B$ is the cumulative background count rate produced by any other sources. $f_i$ is the fraction of the detector that is exposed through the mask to the source position (ie shadowed by a fraction of $1-f_i$). 

In principle, a background-subtracted image of the sky can be reconstructed using this DPI (shadowgram) cast onto the detector, and the known mask shape. There are many methods that can be used to reconstruct the sky image, but in practice, this is typically performed with a balanced cross-correlation of the shadowgram and the mask pattern (the deconvolution array) via a Fast Fourier Transform to speed-up the reconstruction process \citep{codedaperturereview}. This technique is used for BAT, and is operationalized in the ftool \texttt{batfftimage}. 

In the balanced cross-correlation technique the deconvolution array has values of +1 where the path from source to detector is not blocked by a mask tile and -1 where it is blocked. Since it is possible to have a detector be partially blocked by a mask tile the values will actually be $w_i=2f_i-1$, where $f_i$ as previously mentioned is the fraction that the detector is not blocked and $w_i$ is referred to as the ``mask-weight". The choice of a zero summed deconvolution array makes it such that the system point spread function in the reconstructed sky image is localized to the source position, as the cross-correlation for other sky locations with the counts from this source will have an expected value of zero \citep{BalancedCrossCorr}. It also has the added benefit of making the expectation from a diffuse background to be zero, as shown in Eq. \ref{eq:mwcounts}. 

To reconstruct an image of the sky, a cross-correlation of the DPI with the deconvolution array is performed for each sky pixel. The correct deconvolution array for each sky pixel is found quickly by taking the mask pattern array, which has a value of -1 where there is a tile and +1 where there is not, and rebinning it down so that each cell is the same physical size as a detector pixel (5 mm squares down to 4.2 mm squares). The resulting array now has values ranging anywhere from $-1$ to $1$ and gives the $w_i$ values in the deconvolution array for a source directly overhead ($\theta=0$ and IMX, IMY = 0, 0). Then for the remaining sky pixels, the deconvolution array is found by sliding the mask pattern array in the X-Y plane. This gives the correct shadow pattern for a point source at IMX = $-\frac{\Delta x}{L}$, IMY = $-\frac{\Delta y}{L}$, where $\Delta x$ and $\Delta y$ are the displacement of the mask pattern array and $L$ is the distance between the mask and detector array. For each sky pixel with the shifted mask pattern array, the rebinning is performed and the cross-correlation is calculated, giving the reconstructed image pixel value. In the ftool \texttt{batfftimage} this procedure is done with a step size in x and y equal to the detector pixel, making sky pixels every $\sim\frac{4.2 mm}{1 m} = 0.0042$ in IMX and IMY, resulting in $\sim$300,000 sky pixels. The images made by the BAT real-time onboard analyses take this form. To make a finer sampling of the sky image \texttt{batfftimage} then repeats the process a number of times, specified by the oversampling parameters, with the mask positions offset by a fraction of a step in x and y. The default oversampling makes sky pixels spaced $\sim$0.0021 apart in IMX and IMY, which is $\sim$0.12$^{\circ}$ at the center of the field of view. With this oversampling we end up with $\sim1.2$ million sky pixels sampled over the coded FOV.

\begin{figure}[h!]
    \centering
    \includegraphics[width=\textwidth]{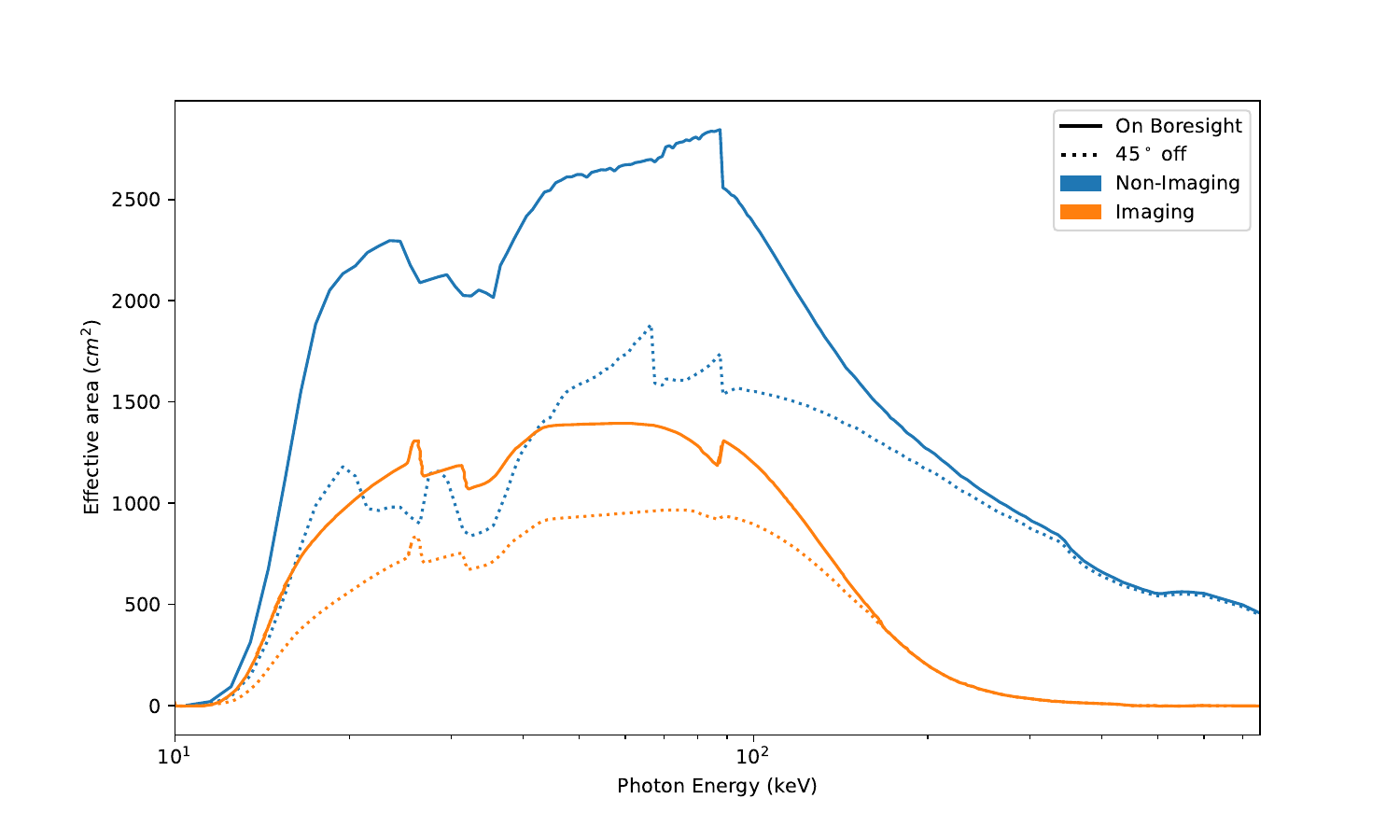}
    \caption{The effective detecting area of the BAT with the conventional coded mask imaging technique, compared to the effective area on the detector in raw counts space. Both are assuming all detectors are active. The overall decreased effective area of imaging is due to a $\sim54\%$ efficiency loss due to the mask-weighting procedure. The energy dependent differences are functions of the transparency of the lead mask tiles, and at high off-axis angles due to the transmission of the shielding and spacecraft components around the BAT instrument.}
    \label{fig:AEffcomparison}
\end{figure}

This imaging technique utilizes the aforementioned `mask-weighting' procedure, whereby each individual detector receives a weight based on the direction of the sky position of interest and the detector position. This weight quantifies the illumination fraction of the detector through the coded mask as $w_i=2f_i-1$, and $w_i$ can thus range from $-1$ to $1$. These are the same weights that make up the deconvolution array. Here we follow \cite{SatoThesis} in deriving the mask weighted counts, from which the sky image is constructed. If we take Eq. \ref{eq:rawcounts} (the counts for a single detector), multiply by its mask weight for a given position on the sky, and then sum over all the detectors in the array (equivalent to the cross-correlation used to reconstruct a sky pixel), we get
\begin{align*}
    R_{mw} &= \sum_i R_i \cdot w_i \nonumber \\
    &= \sum_i \big[ S\cdot f_{\rm trans} \cdot \{f_i+(1-f_i)\cdot f_{Pb}\}\cdot A_{\rm eff} + B  \big] w_i \nonumber \\
    &= S \cdot f_{\rm trans} \cdot A_{\rm eff} \cdot \bigg[ \frac{1}{2}(1-f_{Pb}) \sum_i w_i^2 +\frac{1}{2}(1+f_{Pb})\sum_i w_i \bigg] + B \sum_i w_i
\end{align*}
Half of the detectors will have positive $w_i$ mask weights, and half will have negative. Because the ratio of open to closed elements in the mask is unity, $\sum_i w_i\approx 0$. As such, those terms drop out, and we are left with the mask-weighted count rate
\begin{equation}
    R_{mw} = \frac{1}{2}S\cdot f_{\rm trans}\cdot A_{\rm eff}\cdot (1-f_{Pb}) \sum_i w_i^2.
    \label{eq:mwcounts}
\end{equation}
We see that the mask-weighting procedure has removed the background component! However, this procedure pays a penalty efficiency factor of $\sum_i w_i^2$ which effectively reduces the effective area of imaging. Since this term is dependent on the square of the summed weights, it is determined uniquely by geometry via the detector size, the mask element size, and source distance. For a source at infinity shining on BAT, $\sum_i w_i^2$ divided by the number of active detectors is calculated as 0.54, reducing the effective area proportionally (see Figure \ref{fig:AEffcomparison}). BAT sky images are constructed in these mask-weighted counts, which are a sum over a large number of counts, so a Gaussian approximation to the noise is valid. 

To understand the practical effects of the imaging procedure on the achievable sensitivity, we focus on the signal to noise ratio (SNR) of a source in a reconstructed sky image. Following the sensitivity comparisons by \cite{SkinnerCodedSens}, the image SNR of an \textit{idealized} coded-mask telescope with an open fraction of $\frac{1}{2}$ would be
\begin{equation}
SNR_{\rm ideal} = \frac{\sum_i (S/2) A_{\rm eff}}{\sqrt{\sum_i (S/2) A_{\rm eff} + B}},
\label{eq:idealimagesnr}
\end{equation}
the expected signal counts divided by the square root of the total counts, since in a coded aperture instrument the signal counts function as background for other positions in the sky image. Assuming $f_{\rm trans}$ = 1 and $f_{Pb}$ = 0, BAT's imaging SNR is, at most, 0.73 times the ideal SNR due to the finite resolution of the detectors recording only a blurred image of the mask's shadow (see Eqs. 23 and 25 from \citealt{SkinnerCodedSens}). This penalty of 0.73 in the SNR can equivalently be interpreted as deriving from the 0.54 efficiency factor in the mask-weighted effective area divided by the square root of that same factor in the noise term. This penalty factor that decreases the sensitivity from the ideal case is termed the `coding power' of the instrument. In reality $f_{Pb}\neq0$ so the coding power, and maximum achievable imaging sensitivity, is even further decreased.

\begin{figure}[h!]
    \centering
    \includegraphics{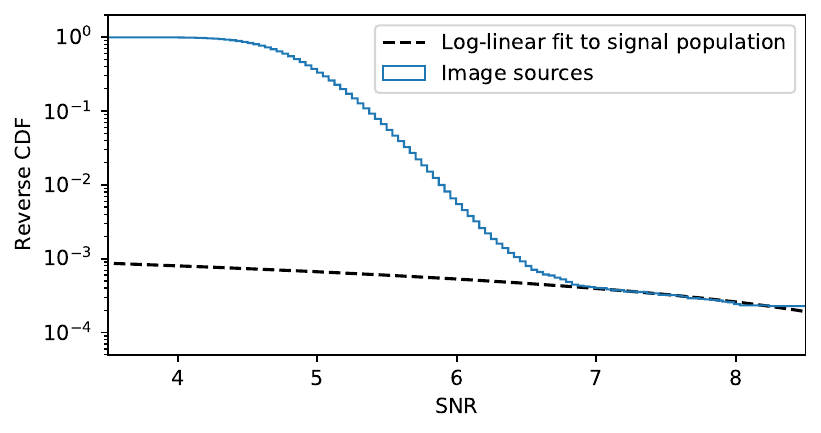}
    \caption{The cumulative distribution of the SNRs of candidate sources in BAT sky images produced with conventional imaging onboard during 2014. At SNR $\sim6.8$ the distribution transitions from signal dominated to noise dominated. A log-linear fit to the real sources with SNR $>7$ shows that the population of real, lower SNR, sources in the data is swamped by the large population of noise sources.}
    \label{fig:ImageBlips}
\end{figure}

Source finding in these sky images is performed with a sliding-cell annulus technique, implemented as the ftool \texttt{batcelldetect}. The combination of the coded image noise and this form of source finding results in a large population of noise peaks found in the images, as seen in Figure \ref{fig:ImageBlips}, which prevents the confident discovery of real, dim, GRBs in the data. While there are 32,768 detector pixels in a DPI, upon projection onto the sky via the imaging technique described above, we end up with $\sim$300,000 (mostly) independent sky pixels, before oversampling.\footnote{For on-boresight directions a BAT image sky pixel is 17 arcmin$^2$, but for lines-of-sight far off-axis the pixel scale can decrease by a factor of $\sim2$.}  Using a source finding routine like that of \texttt{batcelldetect} we have a similar number of uncorrelated trials to the number of sky pixels, given the size of the BAT point-spread function.
With $\sim$300,00 uncorrelated trials, and assuming purely Gaussian statistical noise, the chance of a false source $>5$ sigma in any given sky image is $\sim10$\%. Empirically, the distribution of noise sources in images are found to have a slightly wider spread than the expectation from purely Gaussian noise. This is likely due to 1) the addition of at least half-a-sigma, or more, of systematics from uncertainties in the mask shape, varying detector efficiencies, etc. but also 2) the non-Gaussian noise remaining in the data after, for example, the imperfect imposition of the re-balancing after the mask-weighting procedure.\footnote{The interested reader can reference the discussion in Section 2.1.5 of \cite{TransMon} for analysis of the degree of non-Gaussianity in the significance distribution of noise sources in BAT sky images.} The oversampled images produced by \texttt{batfftimage} give an even higher-weighted SNR distribution due to finer sky pixels.

The mask-weighted imaging procedure has the advantage that it is very computationally inexpensive, lending itself to real-time onboard analyses. However, we have shown that it suffers a sensitivity penalty from the mask-weighting procedure, and results in a large population of noise sources that make discriminating real, weak, sources extremely difficult. In addition, this imaging detection technique fails to use all the relevant information encoded in the TTE data, and naturally rejects counts that reach the detector via paths that do not pass through the mask structure (they receive a mask-weight of 0). It behooves us then to consider whether there may be analysis techniques that escape these limitations, while still enabling the critical localization that imaging provides.

The SNR for a transient event found in the time domain, computed as the sum of the total counts across all detectors and a known background rate sampled from before and after the signal window, will be approximately the same as the ideal image SNR for an on-axis source in the background dominated case (signal counts divided by square root of the total counts). Given this, the SNR in counts space is a factor of at least $\sim \frac{1}{0.73} = 1.37$ higher than the BAT imaging SNR, and properly taking into account $f_{Pb}\neq0$ the relative sensitivity of counts vs imaging gets larger still (as seen in Figure \ref{fig:AEffcomparison}). Hence an analysis looking at just the total counts yields a larger SNR for transient source detection, as it avoids the mask-weighting efficiency penalty, but does not provide any information about the source's location. 
This approach of summing the counts from all detectors ignores the information from the mask, so it follows that an approach that incorporates information about the mask shadow pattern (and more generally the response along each line of sight) may further improve upon the sensitivity of counts-space sensitivity. The mask-weighting procedure does use the mask shadow pattern, but loses effective area from detectors that are partially shadowed (as opposed to the optimal fully shadowed/exposed) making the mask-weights less efficient. Simultaneously, the mask-weighting procedure does not improve upon the noise as it is still a sum over all counts.

There are several other approaches to searching for point source transients with a coded aperture, such as matrix inversion and optimizing, iterative methods. One such iterative method is the maximum likelihood method, which can deal with the count expectation and statistics on a per detector basis instead of single summed value, possibly avoiding the efficiency factor to the effective area and achieving lower noise. The maximum likelihood method has previously been shown to improve sensitivity for coded aperture imaging \citep{OldMLE} and has been applied to the BAT time-integrated survey data \citep{AjelloMLE}. However such approaches have typically ignored more complicated physical effects of the instrument, focusing instead on the purely geometric shadowing (eg \citealt{bayesiancoded}, \citealt{LEGRIlikelihood} and others). In the rest of this manuscript we will demonstrate a unique implementation of a likelihood framework for analyzing BAT event data, that fully exploits the information associated with each individual count and the full physical instrument response, and apply it to a search for transient point sources.

\section{A Likelihood Framework}
\label{LikeSect}
In this section we describe a likelihood-based approach to BAT analysis, that forward models different sky distributions (including GRBs) through the entire instrument response. The resulting model shadowgrams (DPIs) are then compared to the observed ones, in order to determine the most probable sky distribution and, in the particular case of GRB searches, the location of the GRB. In comparison to the standard cross-correlation imaging, this approach has the advantage of fully exploiting the spectral and timing content of the TTE data, is capable of utilizing counts that reach the detector via paths that do not pass through the coded mask, and results in substantially more sensitive searches; at the cost of dramatically increased computation time.



In this analysis, we use the Maximum Likelihood Estimation (MLE) approach, where it is assumed that the set of parameters that maximizes the likelihood are the most probable or the best fit
parameters.
Recalling that the TTE data contains a timestamp, energy measurement, and detector position for each count registered by the detector, we operationalize our likelihood as follows. For the likelihood, the data for a certain time interval is taken and binned by detector and energy. The data input for the likelihood is then $N_{ij}$, which is the number of counts in detector $i$ and energy bin $j$. Since this is a counting experiment the likelihood for a single bin will take the form of a Poisson likelihood of there being a count expectation $\lambda_{ij}$, given the observed counts $N_{ij}$. With some model $M(\bm{\Theta})$ with a set of parameters $\bm{\Theta}$ that give the expected number of counts ${\lambda}_{ij}$ the likelihood for detector $i$, energy bin $j$ is,

\begin{equation}
l_{ij}(M(\bm{\Theta}) \vert N_{ij}) = Pois(N_{ij}; \lambda_{ij}) = \frac{(\lambda_{ij})^{N_{ij}}e^{-\lambda_{ij}}}{N_{ij}!}
\label{PoisLik}
\end{equation}

In this framework we also allow for the model to have some error on the count expectation $\lambda_{ij}$ given the set of parameters $\bm{\Theta}$. As an example, consider a point source model where the source flux is a parameter, there may be some error on the detector's effective area that would result in an error on the count expectation. For now we only allow for Gaussian errors, so the error probability density function (PDF) will take the form of a Normal distribution $\mathcal{N}$. For a model with a Gaussian error $\sigma_{ij}$ on the count expectation and mean count expectation $\bar{\lambda}_{ij}$, the error PDF is,

\begin{equation}
P(\lambda_{ij} \vert M(\bm{\Theta})) = \mathcal{N}(\lambda_{ij}; \bar{\lambda}_{ij}, \sigma_{ij}) = \frac{1}{\sqrt{2\pi \sigma^2_{ij}}} \exp{ \left[  -\frac{(\lambda_{ij} - \bar{\lambda}_{ij})^2}{2\sigma^2_{ij}} \right] }
\label{NormErrorPDF}
\end{equation}

To account for the model error in the likelihood, the Poisson likelihood (Eq. \ref{PoisLik}) is integrated over the error PDF (Eq. \ref{NormErrorPDF}) on $\lambda_{ij}$. The likelihood for detector $i$, energy bin $j$ is then,

\begin{equation}
l_{ij}(M(\bm{\Theta}) \vert N_{ij}) = \int Pois(N_{ij}; \lambda_{ij}) \mathcal{N}(\lambda_{ij}; \bar{\lambda}_{ij}, \sigma_{ij}) d\lambda_{ij}
\end{equation}
The full log-likelihood (LLH) over all bins is then the sum of ${\rm LLH}_{ij}$,
\begin{equation}
LLH(M(\mathbf{\Theta}) \vert \mathbf{N}) = \sum_{ij} log[ l_{ij}(M(\mathbf{\Theta}) \vert N_{ij}) ]
\label{LLHeq}
\end{equation}
where $\mathbf{N}$ is the set of counts $N_{ij}$.

This likelihood framework can be used for several different types of analyses. In this paper we focus on using it to search for GRBs, where the search is described in section \ref{sec:GRBsearch}.

\subsection{Models}

We have already established the form of the likelihood for our general model that generates an expected number of counts $\bar{\lambda}_{ij}$ with error $\sigma_{ij}$. The remaining job of our model is to describe how to go from our parameters $\mathbf{\Theta}$ to $\bar{\lambda}_{ij}$ and $\sigma_{ij}$. This requires a complete understanding of our instrument and all the potential sources of counts. 

Count sources can be divided into two major categories; point sources that create counts from a photon flux originating from a single position in the sky and diffuse sources that do not originate from any specific direction. The significant difference between these classes is how the resultant counts are spatially distributed across the detectors.

\subsubsection{Diffuse Model}

\begin{figure}[htb]
	\centering\includegraphics[width=6 in]{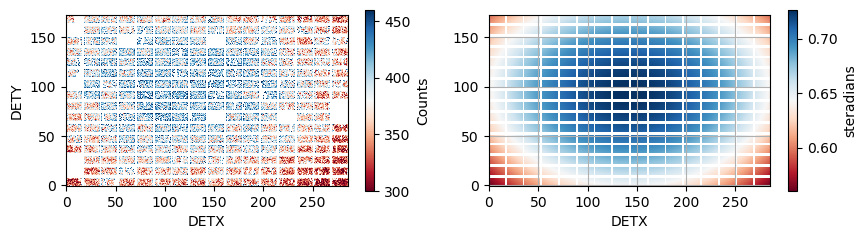}
	\caption{The left plot shows the number of counts in each active detector over an exposure of 1316 s with no very bright sources in the FOV. The right plot shows the solid angle each detector is exposed to through the mask, and is the template used in the diffuse model. }
	\label{SolidAngDPI}
\end{figure}

The largest source of counts for BAT is the cosmic X-ray background (CXB) and next largest diffuse source are cosmic rays interacting with the spacecraft. As pointed out by \cite{AjelloCXB}, the CXB shining through the coded mask causes a spatial pattern across the detectors that can be seen in Figure \ref{SolidAngDPI} with more counts in detectors closer to the center. This is because the detectors closer to the center are exposed to a larger solid angle of unblocked sky. Counts resulting from CXB photons that travel through a mask tile, or other part of the spacecraft, and cosmic ray induced photons will not follow this pattern. To include both of these patterns in the diffuse model, for each energy bin we have two parameters; rate per detector solid angle $\phi_j^b$ and rate per detector $r_j^{b}$. The expected distribution of counts from the diffuse model is

\begin{equation}
\bar{\lambda}_{ij}^{diff} = (\Omega_i\phi_j^b + r_j^b)T
\label{DiffModel}
\end{equation}
where $\Omega_i$ is the unblocked solid angle for detector $i$ and T is the exposure of the observation. So, $\phi^b$ is able to fit the counts that are proportional to $\Omega_i$ and $r^b$ is able to fit the counts that are the same in every detector. 

It is known that the detectors are not identical and have different efficiencies, which are taken into account for very long timescales ($\gtrsim$ months; \citealt{BATsurvey70mon}). This analysis is for much shorter exposures, so these are not taken into account as directly. To  account for these variations a small 4\% error on $\bar{\lambda}_{ij}^{diff}$ is used. A 4\% error should be large enough to account for the detector variance, while still being smaller than the larger systematic uncertainty of the point source response.   


\subsubsection{Point Source Model}

The point source model's parameters are the source's position ($\theta, \phi$ in instrument coordinates) and spectral parameters (normalization and shape). To calculate $\bar{\lambda}_{ij}^{PS}$, we need what's called the detector response matrix (DRM), which describes the detector's effective area and how counts will be distributed in bins of measured energy for a set of incident photon energies $E_\gamma$. The response for a single $E_\gamma$ (a row of the DRM) is

\begin{equation}
R_{j}(E_\gamma) = w_j A_{\rm eff}(E_\gamma)
\label{DRMrow}
\end{equation}
where $w_j$ is the fraction of counts that fall in energy bin $j$ and $\sum_{j}w_j=1$. Then the expected number of counts from a photon spectrum $f(E_\gamma)$ in energy bin $j$ is,
\begin{equation}
\bar{\lambda}_{j} = T \int f(E_\gamma)R_{j}(E_\gamma) dE_\gamma
\end{equation}
For computational reasons the DRM is $R_j$ calculated for a set $E_{\gamma,l}$, so realistically $\bar{\lambda}_{j}$ is computed as,
\begin{equation}
\bar{\lambda}_{j} = T \sum_l DRM_{jl} f(E_\gamma=E_{\gamma,l}) \Delta E_{\gamma,l}
\end{equation}

In section \ref{sec:respmodels} the DRM will be calculated for each detector and we will show how it depends on the source's position. So $\bar{\lambda}_{ij}^{PS}$ is,
\begin{equation}
\bar{\lambda}_{ij}^{PS} = T \sum_l DRM_{ijl}(\theta,\phi) f(E_\gamma=E_{\gamma,l}) \Delta E_{\gamma,l}.
\label{PSModel}
\end{equation}
In the next section and in Section \ref{DetResp} $\sigma_{ij}^{PS}$ will be shown to depend on the details of the DRM computation. 

\section{Instrument Response Modelling}
\label{sec:respmodels}
With our different source models formalized, we now require full instrumental responses to fold modelled source distributions through. Unfortunately, the uncoded response of BAT has not been well characterized to date. We construct new responses for BAT using the \swift\ Mass Model (SwiMM) \citep{SatoThesis}
, which is a model of parts of the \swift\ spacecraft and instruments that can be used with the Geant4 \citep{Geant} toolkit to perform simulations of a particle flux interacting with the model. SwiMM is not a complete model of \swift, it is missing solar panels and many internal components in the spacecraft body. However, it is believed to be complete for everything above the optical bench that the BAT detector plane sits on top of, corresponding to lines of sight at $\theta<90^{\circ}$. Responses for lines of sight coming from below the instrument plane ($\theta>90^{\circ}$) can be calibrated using GUANO-derived data for bursts coming from these directions, which we discuss later. The specific details of the response construction are left for Appendix \ref{DetResp}, and a brief overview is provided here. We publicly release our constructed responses, comprising some $\sim500$ GB, for the benefit of the community. \footnote{ \href{https://zenodo.org/communities/swift-bat}{https://zenodo.org/communities/swift-bat}}

The response can be split into two components, photons that have their first interactions inside a detector (referred to as the direct response) and photons that first interact with some other part of the spacecraft resulting in either a scattered or new photon(s) that then interact in a detector (referred to as the indirect response).  For the former, that means that the photon either made it through a hole in the spacecraft or instrument (whether a mask hole or otherwise), or successfully made it through materials surrounding the detector without being absorbed. This direct response can then be determined through simulation with no spacecraft surrounding the detectors (shown in section \ref{DetsOnly}) then multiplying that by the photon’s transmission probability through the spacecraft to that detector (shown in section \ref{TransSect}). The indirect response is too complicated to partially compute without simulation, so is found through simulation with the entire spacecraft (shown in section \ref{FullCraft}). 

The goal of the simulations is to determine the following: For a flux of photons at a specific energy $E_i$ what is the distribution of deposited energy ($E_{dep}$) and the detector depth where it is deposited? $E_{dep}$ will be at specific energies (lines) for photoelectric interactions and a continuum for Compton interactions.  For each line a 1D distribution of detector depths will be found and for the continuum a 2D distribution will be found of $E_{dep}$ and detector depth. The transmission to each detector for each line of sight are derived from ray traces through the materials in SwiMM, a calculation of distance of intersection by material, and reference to the mass absorption coefficient curves as a function of photon energy for each element the trace passes through.  

Putting together the detector depth distributions, the transmission, and the per-detector mobility-lifetime, the full Detector Response Matrices (DRMs) can be constructed. In Figures \ref{fig:DRM_construct_theta35} and \ref{fig:DRM_construct_theta75} below, we show a diagram of the construction of the responses for a single photon energy of 100.5 keV and at two different sky positions, one within the partially coded FOV and one fully outside the FOV. This procedure is repeated for a variable stepped array of energies from 10 keV to 6 MeV and a grid of positions covering the sky.

Figures \ref{fig:DRM_construct_theta35} and \ref{fig:DRM_construct_theta75} show the construction of a single row in a DRM, which gives the effective area times an array of probabilities that the count will register in each energy bin for a specific incident photon energy. Plots (a) and (d) show these probabilities for two different detectors (one in the middle of a detector group and the other on the edge). Plots (b) and (e) show the effective area for their respective portion of the response for each detector. Plot (c) shows the photon's transmission probability of the photon through the spacecraft to the detector. Plot (f) shows the total effective area for each detector. The DRM row is actually split into the energy bins, but what is shown is the sum over energy bins. Finally, (g) is a view of \swift\ from the direction of the photon with the mask removed so the detectors can be seen. 

\begin{figure}[h!]
    \centering
    \includegraphics[width=0.9\textwidth]{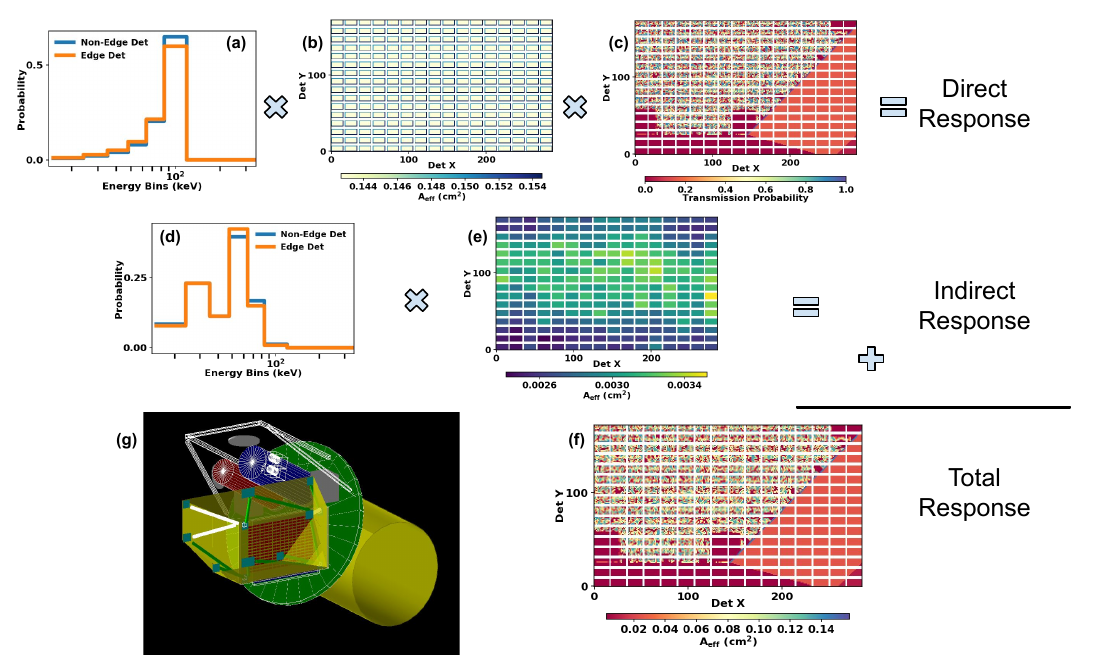}
    \caption{Constructing the response for a 100.5 keV photon simulated at $\theta=35^\circ$ $\phi=30^\circ$, and the view of \swift\ from that line of sight. This position has a $54$\% coding fraction. The portion of the detector array that is coded (photon goes through the mask) can be easily identified in the total response where the effective area is higher and rapidly varying. The uncoded response is determined by the portion of the shield the photon passes through, with higher transmission probabilities where the shield is thinner (towards the mask), or is closer to perpendicular to the shield.}
    \label{fig:DRM_construct_theta35}
\end{figure}
\begin{figure}
    \centering
    \includegraphics[width=0.9\textwidth]{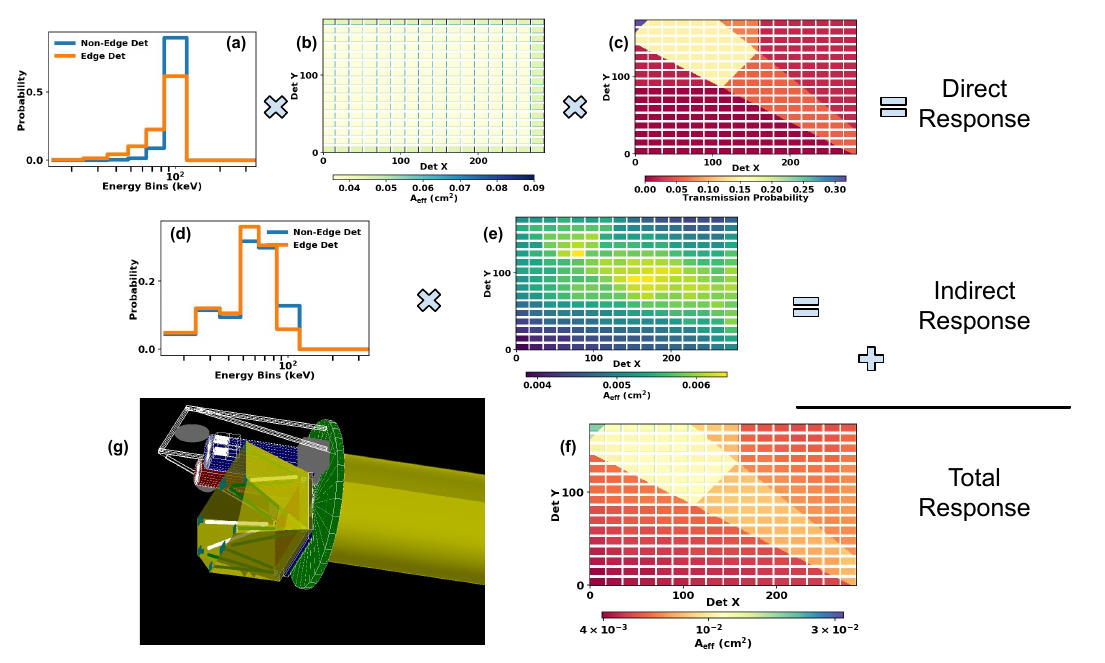}
    \caption{Constructing the response for a 100.5 keV photon simulated at $\theta=75^\circ$ $\phi=30^\circ$, and the view of \swift\ from that line of sight. This position is entirely outside of the coded FOV. From this direction the total effective area mostly depends on the transmission probability. The direct response dominates at high transmission probabilities, and for the detectors at the lowest transmission probabilities the indirect response is dominant. }
    \label{fig:DRM_construct_theta75}
\end{figure}

\begin{figure}[htb]
\centering
\begin{subfloat}
  \centering
  \includegraphics[width=0.49\linewidth]{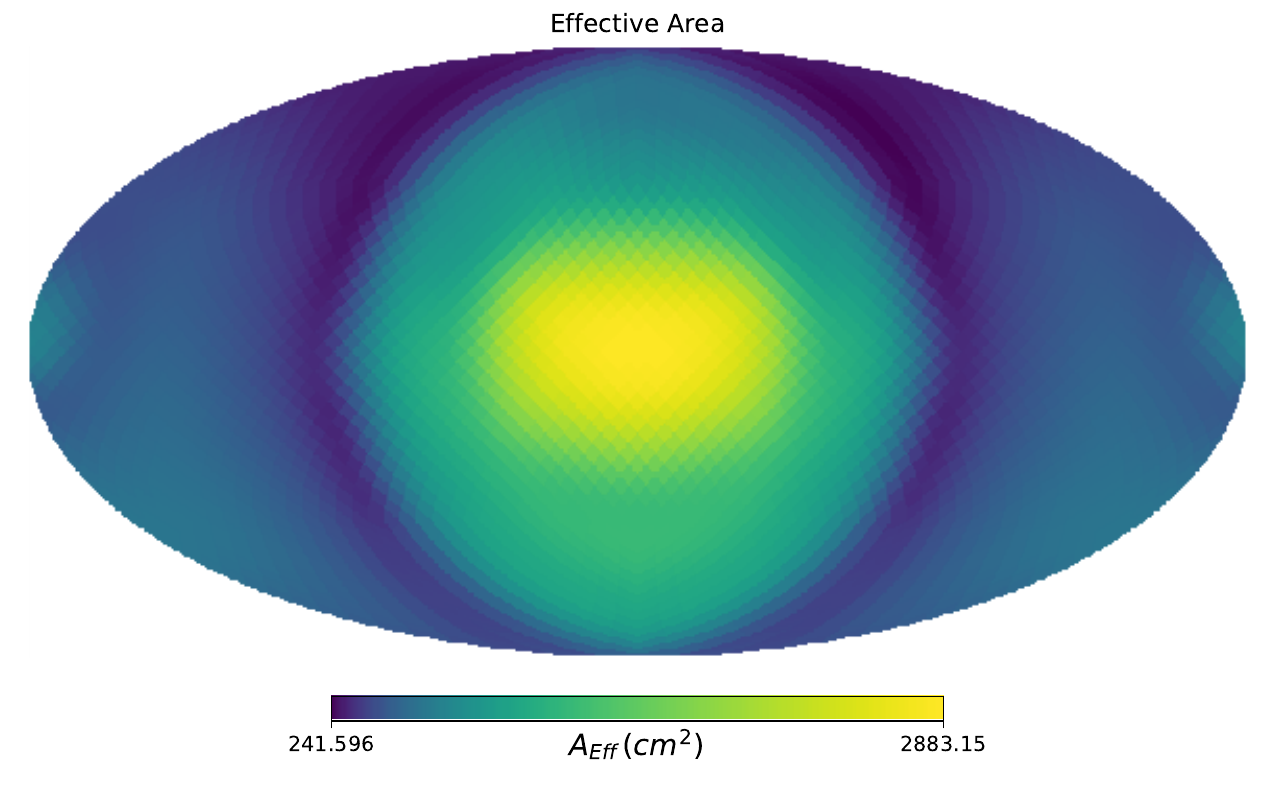}
  \label{fig:AEffMap}
\end{subfloat}%
\begin{subfloat}
  \centering
  \includegraphics[width=0.49\linewidth]{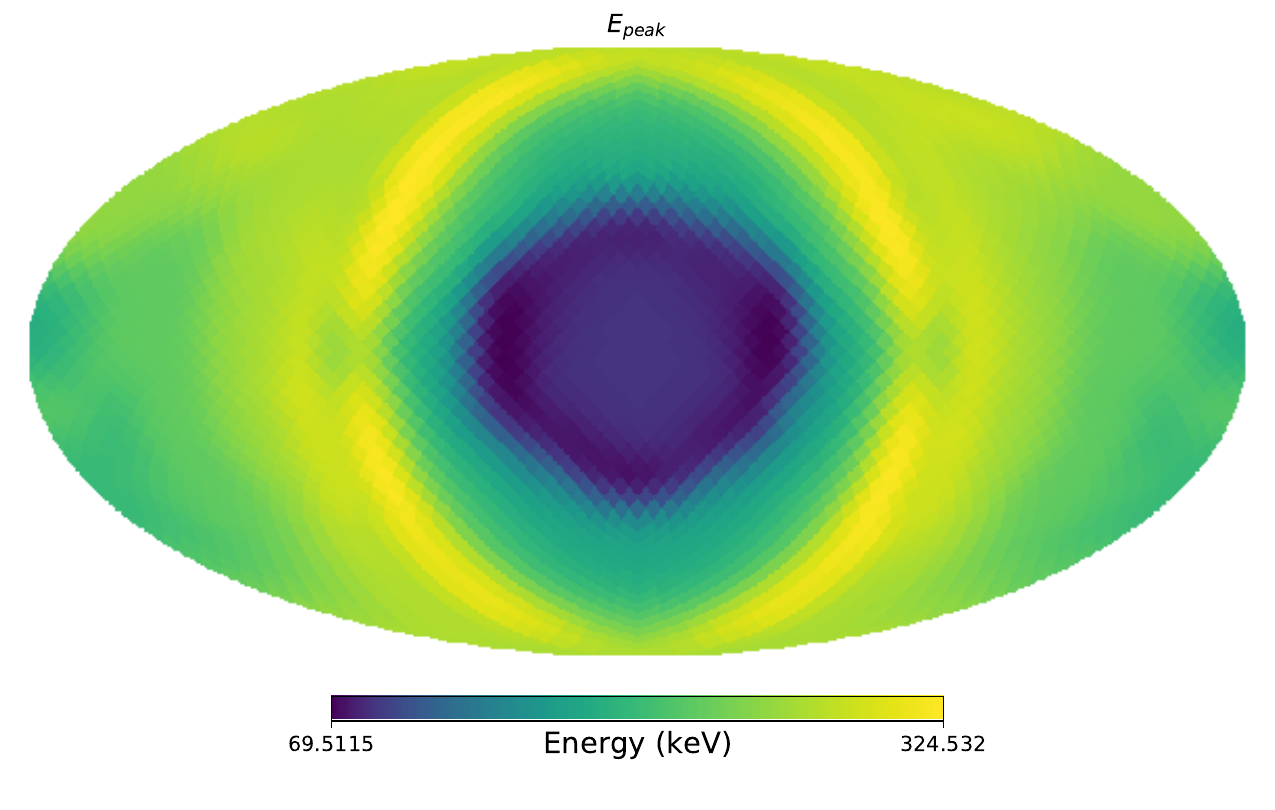}
  \label{fig:Epeakmap}
\end{subfloat}
\caption{Left: Mollweide projection of the peak total effective area (for any photon energy) of BAT on the sky. Right: The photon energy for which the BAT effective area is maximum as shown on the left. }
\label{fig:AeffEpeakMap}
\end{figure}

\begin{figure}
    \centering
    \includegraphics[width=0.6\textwidth]{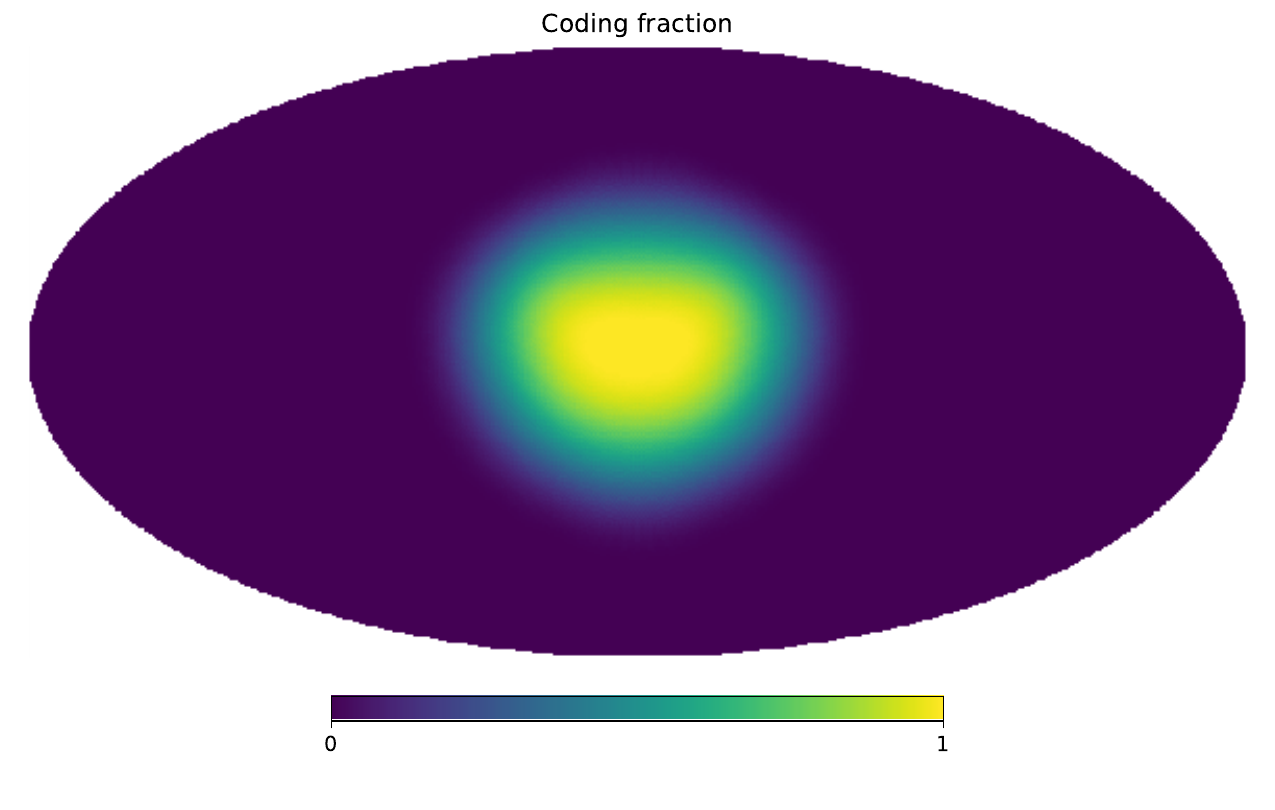}
    \caption{Mollweide projection of the fraction of the BAT array that is illuminated through the coded aperture mask for every given point on the sky. Arcminute localizations can only be determined for sources coming from sky positions with non-zero coding fractions.}
    \label{fig:codingmap}
\end{figure}

In Figure \ref{fig:AeffEpeakMap} we show the peak total effective area for BAT across the entire sky, in detector coordinates, and the $E_{\rm peak}$ at which this effective area is achieved, as derived from our generated responses. As can be seen, BAT retains hundreds of cm$^2$ of effective area across the entire sky, and consequently regularly detects GRBs from outside of its coded field of view (Fig. \ref{fig:codingmap}), but without the ability to localize them to arcminutes.

\section{GRB Search}
\label{sec:GRBsearch}

Using the likelihood method outlined in Section \ref{LikeSect} and the response models developed in Section \ref{sec:respmodels}, we develop a search targeted to find short-duration GRB-like transients. While this search is tuned for short GRBs, it is also capable of recovering the majority of the long GRB population (T90 $\lesssim 60$ seconds). An on-off analysis is used, where there is

\begin{itemize}

\item a signal time window (on-time) over which the search will be performed,

\item a background model fit to data from a nearby time (off-time) that does not overlap the signal time window,

\item a background plus signal model fit to the on-time data by optimizing over the signal parameters and using the background information found from the off-time fit.


\end{itemize}
An on-off analysis is used, since fitting the background model while there is no GRB emission is much easier and leads to more accurate results. This allows us to optimize the signal and background models separately, which lowers cross-contamination between the models and is more computationally efficient. However this does have the downside that it requires near-in-time data suitable for fitting the background. 

As the goal is to determine the evidence for a GRB in the data, once we maximize the LLH for the signal plus background model we need to find its significance. To do this, we use the likelihood ratio test statistic $\Lambda$ to compare this model to a background-only model. For this search we define $\Lambda$ as,


\begin{equation}
\Lambda = -2 [ LLH(\boldsymbol{\Theta}_{\rm bkg}^{\rm off} | \boldsymbol{N}^{\rm on}) - LLH(\boldsymbol{\Theta}_{\rm sig},\boldsymbol{\Theta}_{\rm bkg}^{\rm off}| \boldsymbol{N}^{\rm on}) ]
\end{equation}
where $\boldsymbol{\Theta}_{\rm bkg}^{\rm off}$ is the set of background parameters that were optimized during the fit to off-time data, $\boldsymbol{\Theta}_{\rm sig}$ is the set of signal parameters that maximizes the LLH for the signal plus background model while $\boldsymbol{\Theta}_{\rm bkg}^{\rm off}$ is kept fixed, and $\boldsymbol{N}^{\rm on}$ is the on-time data.

$\Lambda$ is used as our test statistic that quantifies how much our alternative hypothesis of there being a transient point source which was not present during the off-time period is preferred over our null hypothesis that there is no such new point source. For a single selection of data in time that is being tested (on-time data) there is a single $\Lambda$ that describes the preference for the alternative hypothesis. We do not attempt to map $\Lambda$ to a p-value as is often done using Wilks' theorem \citep{Wilks}. In this search, the likelihood does not meet all of the requirements necessary for Wilks' theorem to apply. This is especially true in the parameter space we care most about, where rare noise events can give large values of $\Lambda$ compared to the rest of the noise population. We instead rely on using the distribution of $\Lambda$ found from running the search on a large amount of data to map $\Lambda$ to a false alarm rate, as described in Section \ref{InterpLambda}. A typical convention with the likelihood ratio test statistic is to report the square root, as under certain conditions the square root of the test statistic can give the pre-trial significance in Gaussian sigma units. These conditions are not met here, but we still follow this convention as $\sqrt{\Lambda}$ gives more easy to understand values and approximately follows a similar distribution as the imaging SNR under the null hypothesis. 


\subsection{Off-Time Background}
\label{OffTimeBkg}
The background model should include all sources that could deposit a significant number of counts onto the detector. This is dominated by the diffuse model (Eq. \ref{DiffModel}) as well as any bright sources in the coded field of view. At the short time scales we are interested in there are only a few persistent hard X-ray sources in the sky that are bright enough to create a significant number of counts in $\lesssim$ 1 minute and a few more that are bright enough while in a flaring state. The catalog from the BAT Hard X-ray Transient Monitor \citep{TransMon} is used to check for any known bright sources that are in the coded field of view. 

Many of these known bright sources have complex spectra that are not fit well by a single spectral function. Their spectra are also soft in comparison to GRBs. For both of these reasons, and for computational efficiency, a simplified point source model is used for these sources. We replace the flux model with a simplified counts model and the photons that go through the lead tiles or spacecraft are ignored, as this only becomes significant for spectrally harder sources. The remaining parameters are then the sky position (which is known and fixed) and a rate parameter $r_j^{PS_k}$ for each energy bin, which is the expected counts per second in a completely unshadowed detector. Then, $\lambda_{ij}^{PS_k}$ for point source $k$ would be,

\begin{equation}
\lambda_{ij}^{PS_k} = f_i(\theta_k,\phi_k) r_j^{PS_k} T
\label{KnownPSlam}
\end{equation}
where $f_i$ is the fraction that detector $i$ is unshadowed. 

To avoid adding more parameters and complexity, the background model is assumed to be constant with time over the background window. This is not always a good assumption. The diffuse component from cosmic rays changes with the satellite's position, particularly near the South Atlantic Anomaly (SAA), which is usually fit with a linear function. To account for this, the background window uses times before and after the signal window, centered on the time of interest when possible. This should give the same result as a linear function evaluated at the center time. There are some times where a cubic function would be a better fit, such as when the spacecraft has recently exited the SAA and the total rate stops rapidly dropping and approaches its typical background rate. When a breakdown in the background fitting like this occurs the on-time search results can be unreliable, so breakdowns like this are manually checked for when evaluating a candidate signal. GRBs have been recovered with linear fits to background even when they are not strictly appropriate (e.g. the short GRB 200623A \citep{GRB200623Agcn}, and Section \ref{SectDiscoveries}).

The total background model is then
\begin{equation}
\lambda_{ij}^{\rm bkg} = \lambda_{ij}^{diff} + \sum_k \lambda_{ij}^{PS_k}.
\label{BkgModel}
\end{equation}

\subsection{On-Time Signal Optimization}
\label{OnTimeOpt}
Since we are searching for GRBs, the point source model (Eq. \ref{PSModel}) is the appropriate choice for the signal, where the flux model component is parameterised as exponentially high-energy cutoff power-law spectra, 

\begin{equation}
f(E) = A \big( \frac{E}{E_{piv}}\big)^{-\gamma} \text{exp} \big[ - \frac{(2-\gamma)E}{E_{\rm peak}} \big]
\label{CutOffPlaw}
\end{equation}
with  $E_{piv}$ of 100 keV, the photon powerlaw index $\gamma$ and peak energy $E_{\rm peak}$ are the shape parameters, and $A$ is the normalization. 

For the signal model, the free parameters are position ($\theta, \phi$), $\gamma$, $E_{\rm peak}$, and $A$. The background model parameters are kept fixed to their off-time fit values. To find the maximum LLH we pick a prior grid of reasonable ($\gamma$, $E_{\rm peak}$) values and a grid of position points that cover the sky. Then, for each spectral point at each position the negative LLH (nLLH) is numerically minimized with respect to the normalization $A$. 

With the maximum LLH found for the signal plus background model, to calculate $\Lambda$ we now just need the LLH for the background-only model, which can be found by setting $A$ = 0.

\subsection{Example Search}

In this section and in Section \ref{sec:SearchPipe}, we use TTE data from Swift/BAT triggered short GRB 180805B \citep{GRB180805Bgcn} as an example GRB detected inside the coded FOV. We performed a search using a 0.512 s duration of data near the peak of the emission, using a grid of positions spaced $\approx$0.002 in image coordinates across the coded FOV (the default spacing of BAT sky image pixels). We found a $\sqrt{\Lambda}$ of 32, and the $\Delta$LLH from the maximum LLH position is plotted in Figure \ref{DloglEx}. The left plot of Figure \ref{DloglEx} shows a histogram of 2$\Delta$LLH for every position in the coded FOV, with a distinction between positions that are near the GRB's true position (within two times the PSF FWHM) and every other position labeled as ``away from GRB". The right plot shows the $\Delta$LLH landscape zoomed in on the GRB's position that can be seen, confidently detected at the center of the plot, along with a few degrees of the ``away from GRB" region visible. The largest value of $\Delta$LLH in the ``away from GRB" region is $\approx$110, this value is later referred to as $\Delta LLH_{\rm peak}$. 

\subsection{Localization}

\begin{figure}[tb]
	\centering\includegraphics[width=6 in]{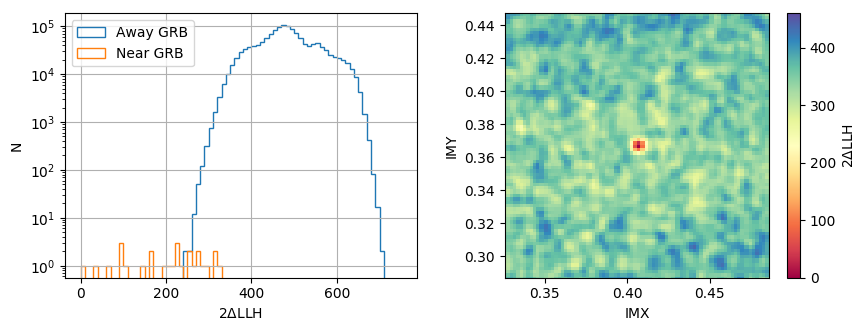}
	\caption{These two plots show the $\Delta$LLH from the max LLH position for GRB 180805B. The left plot shows a histogram of the 2$\Delta$LLH values for each pixel, separated into the few pixels within $\approx$ two times the PSF of the GRB position and then the other pixels. The right plot shows the 2$\Delta$LLH at the pixels in a square around the GRB position, where the GRB position is apparent from the strong peak in LLH. }
	\label{DloglEx}
\end{figure}

Localization in a likelihood analysis is the same as finding confidence intervals for your spatial coordinates. With BAT this gets complicated as the response inside the coded field of view changes very quickly. The LLH landscape is characterized by peaks and troughs on the average distance scale of the PSF (FWHM $\approx$ 22.5 arcmin, 0.006 in image coordinates). The right panel in Figure \ref{DloglEx} shows this messy landscape. For this example the burst is bright enough that there is a very large difference in LLH between the burst location and anywhere else outside of the burst's peak, so the burst location is determined very confidently. For lower significance bursts this will not always be the case. If we want to determine an arcminute scale localization, the most important value is the $\Delta$LLH between the max LLH peak and the separate peak that has the next highest LLH. If those two LLH values are close, then it cannot be confidently said that the burst is localized to a single few arcminute circle. This value will be referred to as $\Delta LLH_{\rm peak}$.

\begin{figure}[tb]
	\centering\includegraphics[width=6 in]{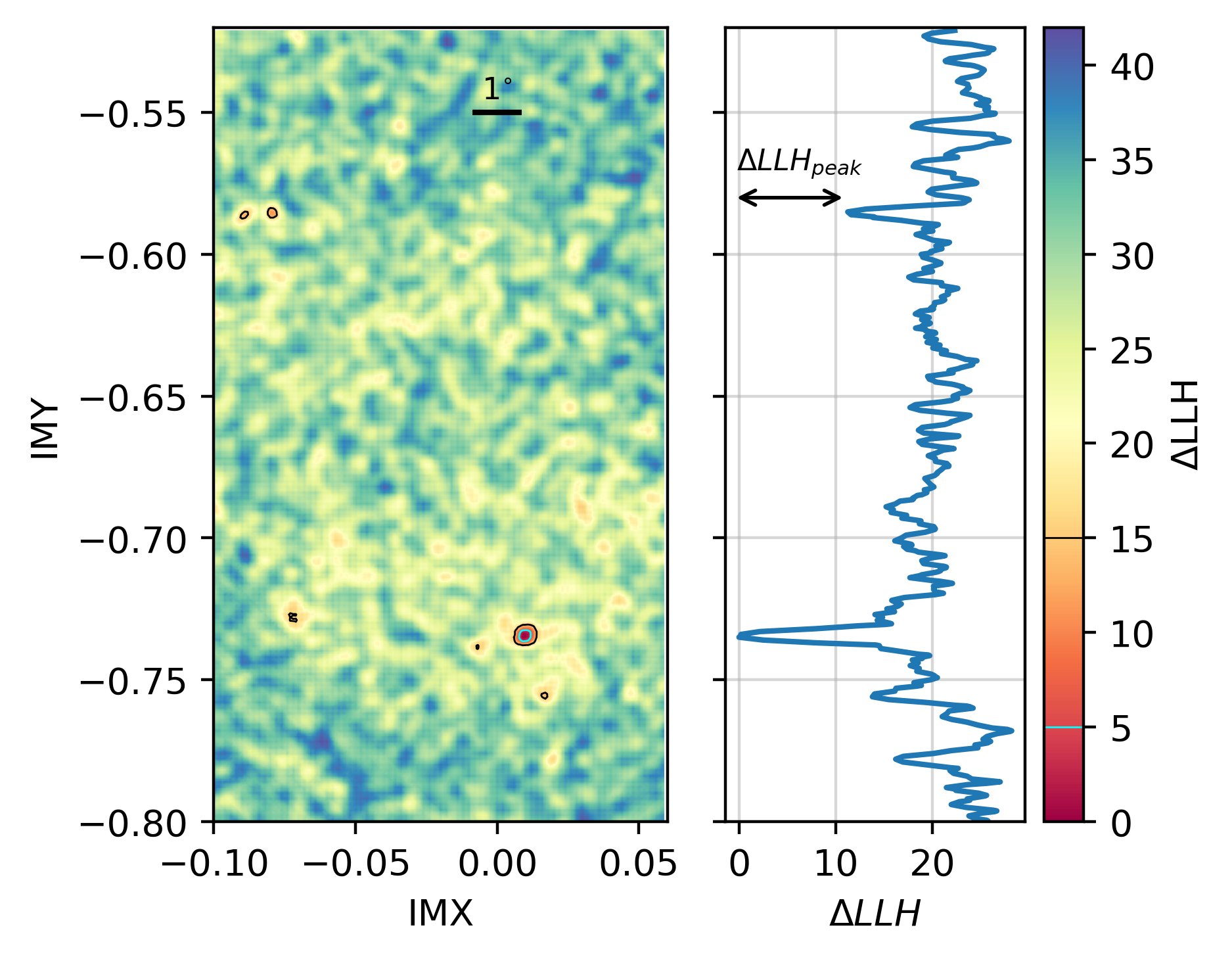}
	\caption{This figure shows the $\Delta$LLH from the max LLH position for GRB 220418B in a region of the field of view that contains both the GRB position and the next most likely LLH peak. The left plot shows the $\Delta$LLH values for each pixel, along with contours at $\Delta$LLH 5 and 15. The GRB position can be seen within the $\Delta$LLH = 5 contour at imx, imy = (0.009, -0.735). The second most likely position is at imx, imy = (-0.079, -0.585), and is one of a few peaks within a $\Delta$LLH = 15 contour. There is also a black horizontal line showing the scale of 1$^{\circ}$ on the sky. The right plot shows the minimum $\Delta$LLH value at each IMY from the left plot. This plot makes it easier to visualize $\Delta LLH_{\rm peak}$, which is marked with an arrow.  }
	\label{ContDloglEx}
\end{figure}

Figure \ref{ContDloglEx} shows an example of a much more weakly localized burst. GRB 220418B was localized in low-latency by NITRATES with a $\sqrt{\Lambda}$ of 10 and $\Delta LLH_{\rm peak}$ of 12. The left plot in Figure \ref{ContDloglEx} is similar to the right plot in Figure \ref{DloglEx}. The GRB's position is right next the maximum LLH position at imx, imy = (0.009, -0.735), and is within the only $\Delta$LLH = 5  contour. The next most likely, separate peak is 7.4$^{\circ}$ away at imx, imy = (-0.079, -0.585). The difference in LLH between these two peaks can be seen well in the right plot of Figure \ref{ContDloglEx}, which shows the $\Delta$LLH profiled over IMX so that it can be viewed as a 1D distribution. The LLH at the GRB is definitely an outlier, but not nearly as extreme as it was for GRB 180805B. Since we can not rely on Wilks' theorem, we need to find at what threshold value of $\Delta LLH_{\rm peak}$ we can state that the GRB's true position is within this LLH peak. We do this in Section \ref{SectDiscoveries} by looking at the empirical distributions of $LLH_{\rm peak}$ values from noise peaks and peaks from actual GRBs. We found the threshold for confident localizations to be at $\Delta LLH_{\rm peak}$ $\approx$ 10, which puts GRB 220418B just above this threshold. 

Having a large enough $\Delta LLH_{\rm peak}$ value allows us to say that the GRB's true position is at least within a PSF scale of the maximum LLH position. To go a step further and find the size of the error contour of this localization would require a way to go from $\Delta$LLH values to confidence levels. This is currently a work in progress. In the mean time we either report a conservative error of $\approx$ 0.1$^{\circ}$ around the maximum LLH position, or use \texttt{batcelldetect} if we're able to recover the source position through conventional imaging. 

\begin{figure}[tb]
	\centering\includegraphics[width=6 in]{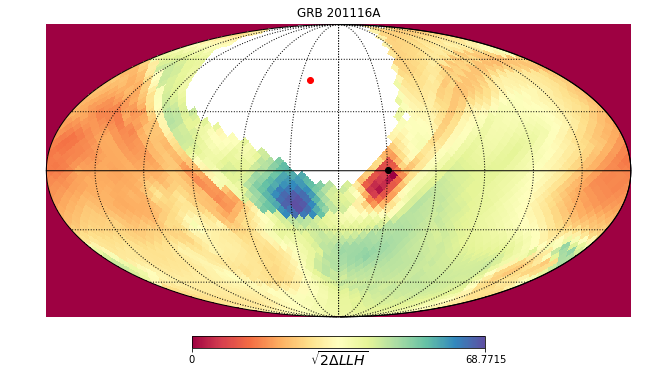}
	\caption{A HEALPix map, with an NSide resolution of 16, of the $\Delta$LLH from the maximum LLH position outside the FOV for GRB 201116A. The actual plotted value is $\sqrt{2\Delta LLH}$ for visualization purposes. The coded FOV is the blank white portion of the map with the red dot showing the FOV center, and the black dot shows the GRB's true position as found by other detectors. The region of low $\Delta LLH$ centered around the GRB's position shows that the analysis did a good job of finding the GRB's position, despite it originating from outside the FOV.}
	\label{BrightOFOVmap}
\end{figure}

Outside the coded field of view (OFOV) this problem does not exist. The response changes much more slowly with position, which makes the parameter space much easier to search, but also makes localizations anywhere close to an arcminute scale impossible. The current errors and limited calibration for the OFOV responses make any type of localization almost impossible except for very bright GRBs and possibly some cases where the flux is well measured by another instrument. Figure \ref{BrightOFOVmap} shows an example of a very bright OFOV burst that was able to be accurately localized by NITRATES, but these are rare. Despite this, this represents the first time that any analysis has independently localized a GRB coming from outside of the BAT FOV, and the ability to distinguish in-vs-out of FOV bursts is critical to confidently detecting and localizing weak bursts.

For the majority of bursts, the important metric for the OFOV positions is the difference between the max LLH inside the FOV (IFOV) and the max LLH outside the FOV; as a test of whether the GRB originates from inside or outside of the coded FOV. When the out of FOV LLH is larger it is very likely that the burst came from somewhere out of the FOV. When the IFOV LLH is only slightly larger it is inconclusive. Inside the coded FOV is a smaller sky area, but it has many more effective trials due to the rapidly changing response, so there is a higher chance of getting a higher LLH. For GRB 201116A in Figure \ref{BrightOFOVmap}, the $max(LLH_{in})$ - $max(LLH_{out})$ is $\approx$ -500, so very confidently determined to originate from outside the FOV. This value will be referred to as $\Delta LLH_{out}$.

\section{Targeted GRB Search Pipeline}
\label{sec:SearchPipe}
A complete pipeline was constructed to be capable of running the likelihood search automatically in response to an external alert. This search is currently triggered by several real-time alert streams that also trigger GUANO event data dumps. A GCN notice, and email, listener running on the Astrophysical Multimessenger Observatory Network (AMON) \citep{AMONpaper} infrastructure orchestrates the search: it runs the analysis manager for the real-time pipeline; gathering data, submitting cluster jobs, and gathering results. The computationally costly parts of the pipeline are run as cluster jobs submitted to Penn State's Roar\footnote{\url{https://www.icds.psu.edu/computing-services/roar-user-guide/}} computing cluster. 

The pipeline runs a targeted search centered around an externally provided trigger time $t_0$ searching at times up to $\pm$ 20s. Inside that search window GRB signal durations of 0.128 s, 0.256 s, 0.512 s, 1.024 s, 2.048 s, 4.096 s, 8.192 s, and 16.384 s are tested at time steps of a quarter of the duration through the search window, following the sliding setup for the GBM targeted search \citep{GBMtarg19}. For example, the possible signal start times for the 1.024 s duration signal tests would be $t_0$ - 20 s, $t_0$ - 19.744 s, $t_0$ - 19.488 s, and so on until $t_0$ + 20s. In practice this is an unrealistic number of possible time windows to run the analysis for, so they're significantly cut down before the likelihood analysis is run, see section \ref{FindingSeeds}. The current iteration of the search uses 9 energy bins ranging from a minimum energy of 15 keV to a maximum of 350 keV. 

\subsection{Data Preparation}

For the real-time search data, scripts on the AMON servers periodically check for and download new data from the \swift's quicklook data website\footnote{\url{https://swift.gsfc.nasa.gov/sdc/ql/}}. Any new BAT TTE data file found is downloaded along with new attitude and enabled detector files, so the data is ready to be used and the satellite pointing history, as well as which detectors are enabled, can be tracked. When the data around a targeted search time becomes available, the TTE data is filtered so that any event with a bad flag or energy outside the desired range (including events without an energy measurement) is removed. The TTE data file includes a table with the good time interval (GTI), which tells us when the data collection happened and if there were any breaks in that time from detector issues. Using the attitude files, any times when the spacecraft was slewing are found and removed from the GTI. To find any `hot' or `cold' detectors a similar, albeit bespoke, process to the one used in the ftool \texttt{bathotpix} is run. In this process, the total number of counts in each detector is compared to the rest of the detectors to see if it is an outlier (low or high) compared to the expectation given a Poisson distribution of counts. Any detector found to be an outlier is masked and not used for the analysis. This removes any detectors that are noisy (hot) or for some reason registering only some or no counts (cold). There are also transient noise events (glitches), that can cause a large number of counts in a short amount of time (usually around 10 ms) in one or many detectors. These glitches tend to only cause counts at lower energies, which helps differentiating them from actual bursts. It is very important to catch and screen these glitches as they can create a high $\Lambda$ value in the likelihood analysis at short exposures, since they are not present in the background model. 

\begin{figure}[htb]
	\centering\includegraphics[width=6 in]{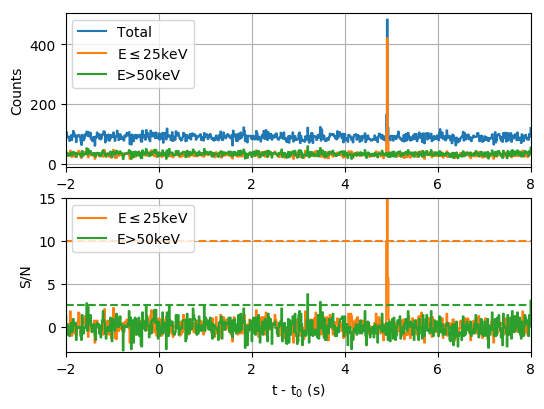}
	\caption{These two plots show an example of a glitch that occurred around the LVC GW candidate S190930t \citep{s190930tgcn}. The top plot shows the lightcurve in 16 ms bins, split into a low-energy (measured energy $\leq$ 25 keV) and a high-energy (measured energy $>$ 50 keV) subsets, as well as the summed lightcurve. The bottom plot has the same lightcurve bins and subsets, but the SNR from Equation \ref{GlitchSNR} is plotted. The dashed lines indicate the thresholds for registering as a glitch, low-energy SNR $> 10$, while high-energy SNR $< 2.5$.}
	\label{Glitch}
\end{figure}

To find glitches that create counts in a large number of detectors, a low-energy light curve is compared to a high-energy light curve with small time bins to see if there are any fast rate spikes at low energies that are not seen at higher energies. To do this, the events with energy $\leq$ 25 keV and events with energy $>$ 50 keV are binned into separate light curves with bins of 16 ms. For both light curves the SNR for each time bin is calculated by finding the mean, $\bar{N}$ and standard deviation of the bin counts, $\sigma^2$. Then, the SNR in the $i$th bin is,
\begin{equation}
SNR_i = \frac{N_i - \bar{N}}{\sigma^2}
\label{GlitchSNR}
\end{equation}
where $N_i$ is the number of counts in the $i$th bin. Any time bin that has a low-energy SNR $>$ 10, while the high-energy SNR is $<$ 2.5 is marked as a glitch and a 32 ms time interval around that bin is removed from the GTI. An example of this type of glitch can be seen in Figure \ref{Glitch}, where there is a large rate spike in a single 16 ms bin, but confined to energies $<$ 25 keV. 

To find glitches that only affect one to a few detectors at a time, another light curve with 16 ms bins is created, but this time for each individual detector and for energies $<$ 50 keV. Any detector that has more than 10 counts in any time bin is added to the list of masked detectors. The light curve ignores times not in a GTI, so this doesn't remove detectors that only had a glitch during times found in the first glitch detection method. 

The astrophysical signal that most closely resembles these glitches are bursts from soft gamma repeaters (SGRs). There have been many SGR bursts inside of BAT's coded FOV while this search has been running during 2020 and 2021 and thus far none of them have been flagged as a glitch, giving confidence that we are not discarding SGR-like astrophysical signals.

Transient noise events can also be caused by high-energy cosmic rays interacting with the spacecraft. Cosmic rays are constantly bombarding \swift\ creating many of the background events registered by the detectors, but when a cosmic ray of sufficient energy interacts with the craft it can create a shower of enough gamma-rays to register hundreds or more events across the detector array. The events caused by a single high-energy cosmic ray tend to have higher measured energies than typical background events and are separated in time by a margin smaller than BAT's timing resolution ($\approx 100 \mu s$). To identify events from high-energy cosmic rays a lightcurve  with bins of 50 $\mu s$ is made using events with energy $>$ 50 keV. Any bin with $>40$ counts and $>10$  times the average counts at $\pm$ 1 s is flagged as a cosmic ray hit. These criteria eliminate any rate spikes that can cause an artificially high and significant $\Lambda$ value at the smallest duration used in the search (0.128 s), while still being under the count rate measured during some of the brightest GRBs observed by BAT. 

The combination of these data cleaning processes successfully mitigates contamination of the data from most noise events, while leaving real signals unperturbed in the data. 

\subsection{Background Estimation}

To fit the off-time background model, the process outlined in Section \ref{OffTimeBkg} is used. An initial fit is performed using all the TTE data outside the signal search window and inside the GTI. TTE data files from GUANO are usually either 90 s or 200 s in duration, so this typically yields at least 50 s of data for the background fit. The initial fit is performed using all of the known bright sources that are at a partial coding fraction of at least 5\%, and the diffuse model. Each known source is checked to see if it is being significantly detected, by checking the $\Delta LLH$ from the best fit parameters as the rates for each source are set to 0 one-by-one. Each source with a $\sqrt{2\Delta LLH}<$ 7 is removed from the model. The fit is repeated until there are no known sources remaining, or all of the remaining sources pass the significance cut. With the list of known sources in the model set, the final fits are performed at a variety of time windows that are centered on times from the beginning to the end of the search window, with a step size of 1 s. The time windows contain data from $\pm$40 s around the center time, with data from $\pm$10 s removed. The on-time analysis will use the best fit parameters from the off-time segment with center time closest to the center of the on-time's search window. 

\subsection{Finding Time and Position Seeds}
\label{FindingSeeds}

As previously discussed, it is computationally unrealistic to run the likelihood analysis on every single time bin. In addition, the spatial parameter space is huge with a PSF of $\approx$ 1/3 degree and a coded FOV of $\approx$ 7,500 deg$^2$. To be able to complete the search within a reasonable amount of time ($\lesssim$ a few hours, to allow for rapid discovery and followup for afterglow searches), a two-step seeding process is used to first significantly reduce the number of time bins, and then to both reduce the fraction of the FOV to analyze as well as a further cut on the number of time bins. The first step is a simple rates analysis, described in section \ref{FullRatesSect} run on all the possible time bins. Then, a more complex analysis, described in section \ref{SplitRatesSect} is run at each time bin that passed the cut in the first step. 

\subsubsection{Summed Detector Rates Analysis}
\label{FullRatesSect}

\begin{figure}[htb]
	\centering\includegraphics[width=6 in]{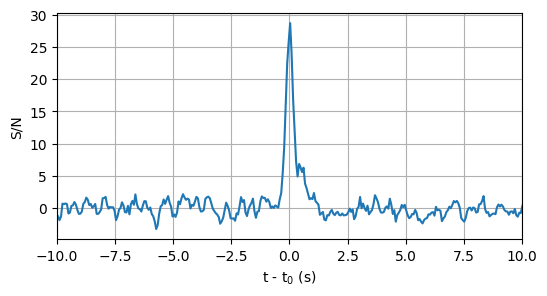}
	\caption{The GRB 180805B full rates analysis SNR results for time bins with a duration of 0.256 s. }
	\label{FullRatesSnrGRB180805B}
\end{figure}

To quickly find time bins of interest, each time bin is examined for excesses over the background for the total counts across all detectors and energy bins. To find the background expectation, at each time with 1 s steps in the search window a linear function is fit to the counts binned into 0.256 s bins using bins that are within 30 s, but are more than 10 s away from the fit time. The y-intercept of the linear function is placed at the fit time and each bin with counts more than 4 standard deviations away from the mean are removed from the fit, so that signal from a burst does not contaminate the background fit. The excess significance for each analysis time bin is calculated as a signal to noise ratio, SNR$^{\rm rates}$. For the Poisson distribution, as the number of expected counts, $\lambda$, becomes large the distribution approaches a Normal distribution with $\mu = \lambda$ and $\sigma^2 = \lambda$. The SNR is calculated using the background subtracted counts as the signal and $\sigma$ added in quadrature with the background fit error, $\sigma_{\rm bkg}$ as the noise. SNR$^{\rm rates}_i$ for time bin $i$ is then,

\begin{equation}
SNR^{\rm rates}_i = \frac{N_i - N_{\rm bkg}}{\sqrt{N_{\rm bkg}^2 + \sigma_{\rm bkg}^2}}
\end{equation}
where $N_{\rm bkg}$ is the expected number of background counts. Figure \ref{FullRatesSnrGRB180805B} shows SNR$^{\rm rates}$ for the 0.256 s duration time bins around GRB 180805B. 

Time bin seeds for the LLH analysis are picked by sorting the rates results for each signal test duration and finding any SNR$^{\rm rates}$ above a certain threshold. Then, any time bin of that duration that passes the threshold is kept as a time seed as long as its SNR$^{\rm rates}$ is at least 0.75 times the largest SNR$^{\rm rates}$ value found near it in time (within $\pm$2 times the duration size). The SNR$^{\rm rates}$ threshold is: 2 for any duration $> 0.256$ s, 2.25 for 0.256 s durations, and 2.5 for 0.128 s durations. These thresholds are very low for a prospective signal, but are able to cut $\approx$95\% of time bins with no signal.

\subsubsection{Split Detector Rates Analysis}
\label{SplitRatesSect}

In order to find position seeds an analysis that depends strongly on source position is needed. The analysis also needs to be fast, so having sufficient counts in each bin to use the Normal approximation of the Poisson likelihood is very beneficial. A good way to meet these criteria is to split the detectors into two groups with the largest difference in average response, the coded detectors (the photon path goes through the mask) and uncoded detectors. 

The goal of this analysis is to find the time bins and position that would yield the largest values of $\Lambda$ in the full likelihood analysis, so this analysis should be as similar as possible. The same energy bins, background parameters, and spectral form are used. The analysis is performed over a grid of points in the tangential plane coordinates (imx, imy). For each position point on the sky, the set of detectors constituting each group (coded and uncoded) is found. See Figure \ref{fig:DRM_construct_theta35}, panel f, for a clear example of the difference in coded vs uncoded detector groups for a given point on the sky. The background expectation, $N^{\rm bkg}_{ij}$ for the two groups ($i$ indexes the group) is found by taking the background model and summing the expected counts for the detectors in each group for each energy bin. Similarly, to get the signal expectation for a given set of spectral parameters, the point source model is used to find the expected counts per energy bin for each group, $N^{\rm sig}_{ij}$. Since the full point source model is more costly to use than desired for this fast seeding analysis, for each position point a lookup table for each group is constructed for $N^{\rm sig}_{ij}$ at a grid of $E_{\rm peak}$ and $\gamma$ values with the spectral normalization, $A$ = 1. In the analysis $N^{\rm sig}_{ij}$ is found for specific values of $E_{\rm peak}$ and $\gamma$ by using a spline interpolation of the lookup tables and then multiplying by $A$. Instead of propagating the point source model errors, the coded error is a flat 5\%, and 10\% for the uncoded error. This is a slight increase from the BAT team's derived 4\% error for the mask-weighted spectra, and for the uncoded error we conservatively double the coded error as calibrations are still a work in progress. The log-likelihood, ${\rm LLH}^{\rm split}$ for a single position and time bin can be written as,

\begin{equation}
\begin{split}
LLH^{\rm split} &= \sum_j \sum_i ln \big[ \mathcal{N}(N_{ij}; \lambda_{ij}, \sigma_{ij} ) \big] \\
\text{where:} \\
\lambda_{ij} &= N^{\rm sig}_{ij} + N^{\rm bkg}_{ij} \\
\sigma^2_{ij} &= N^{\rm sig}_{ij} + N^{\rm bkg}_{ij} + (\sigma^{\rm bkg}_{ij})^2 + (\sigma^{\rm sig}_{ij})^2 \\
\end{split}
\end{equation}

One issue that arises from this method of splitting the data, is that as sky positions approach the edge or center of the FOV the number of detectors for one group naturally goes to 0, as the detector plane becomes fully coded or uncoded. To remedy this, whenever one of the groups has fewer than 100 detectors that group is not used in the likelihood analysis. 

\begin{figure}[htb]
\centering
\begin{subfloat}
  \centering
  \includegraphics[width=0.49\linewidth]{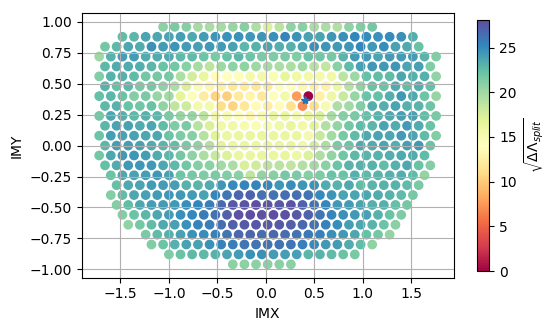}
\end{subfloat}%
\begin{subfloat}
  \centering
  \includegraphics[width=0.49\linewidth]{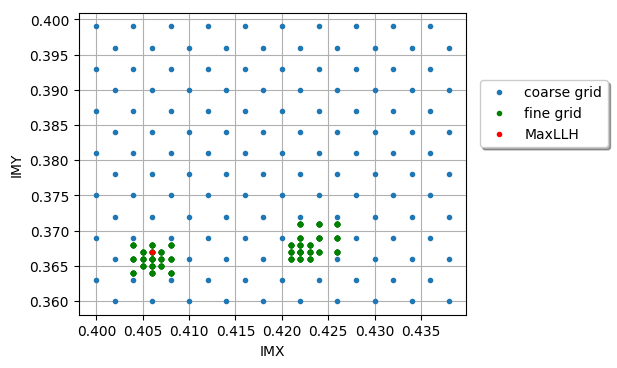}
  \label{GridScan}
\end{subfloat}
\caption{Left: The GRB 180805B split detector rates analysis results for the 0.256 s time bin that starts at the trigger time. The star indicates the true GRB position. Right: The coarse grid shows the initial scan and the fine grid have their LLH maximized during the peak scanning step. The maximum LLH position, which is also closest to the true GRB position, is shown in red.}
\label{SplitRatesIMxyGRB180805B}
\end{figure}


At each sky position ${\rm LLH}^{\rm split}$ is numerically maximized over the flux parameters. The significance is estimated the same way as the full likelihood analysis, with the test-statistic
$\Lambda$ comparing the best fit background-plus-signal model to a background-only model ($N^{\rm sig}$=0). The maximum $\Lambda$ in the time bin provides information on the likelihood of a burst at this time, but we also want information on where the burst might be originating from, to derive position seeds for the full analysis. The ${\rm LLH}^{\rm split}$ values at separate positions cannot be directly compared, since the data at each position is `changed' because of the different group binning, especially in cases where there is only one detector group. However, the $\Lambda$ values for each position can be directly compared in a similar way that the LLH values would be if the data didn't change. Figure \ref{SplitRatesIMxyGRB180805B} shows an example of this with GRB 180805B, where $\Delta \Lambda_{\rm split}$ is the maximum $\Lambda_{\rm split}$ value from the analysis minus the $\Lambda_{\rm split}$ value at that position. It shows that the split-rates analysis can quickly narrow down the portion of the FOV most likely to host the GRB. 

Time seeds are chosen using the maximum $\sqrt{\Lambda_{\rm split}}$ values from each time bin analyzed. An initial cut is done by removing times with $\sqrt{\Lambda_{\rm split}}<$  4.5. Then, any $\sqrt{\Lambda_{\rm split}}$ values less than 0.7 times the maximum value are removed. If the remaining time bins overlap with any other time bins, then only the top 6 are kept and any time with $\sqrt{\Lambda_{\rm split}}$ less than 0.8 times the value for a bin it overlaps is removed. Finally, only the top 8 remaining time seeds are kept.

For position seeds inside the coded FOV, the position space is cut up into squares in imx, imy coordinates with width 0.04 (see Fig. \ref{SplitRatesIMxyGRB180805B}). For each time seed remaining, $\Delta \Lambda_{\rm split}$ is found for each position point. A cut value of $\Delta \Lambda_{\rm split}$ is determined by finding the 7.5th and 30th percentile values of $\Delta \Lambda_{\rm split}$, and the cut value is chosen with the following logic, 
\begin{itemize}
	\item If 7.5th percentile $>$ 24; Then cut value = 7.5th percentile + 1
	\item If 7.5th percentile $<$ 24 $<$ 30th percentile; Then cut value = 24 + 1
	\item If 30th percentile $<$ 24; Then cut value = 30th percentile + 1
\end{itemize}
where that 1 is added for some wiggle room. For each square the minimum value of $\Delta \Lambda_{\rm split}$ inside that square is found using linear interpolation, and is kept as a position seed if greater than the cut value. Then, each neighbor to a square passing the cut is also kept as a position seed. 


This analysis is also performed for points outside the FOV, but it very rarely gives the maximum values of $\sqrt{\Lambda_{\rm split}}$ as it has just one (uncoded) detector group. For the same reason it generally doesn't provide any usable localization information, so for each time seed the whole OFOV sky is included as position seeds. This does not significantly add to the computational load as the number of OFOV points is much smaller than the number of IFOV position points to include, even for the smallest possible number of seeds. The IFOV position points are included out to a partial coding fraction of 0.5\%. Near the edge of the FOV, where the number of coded detectors becomes very small, it gets progressively harder to find the correct position seeds. 

An additional step using conventional imaging to find position seeds could also be used to avoid positions with very low or negative image SNRs and to pick high SNR positions that did not quite make the $\Delta \Lambda_{\rm split}$ cut. However, this would add a major dependency to the pipeline, since everything else is Python native and does not utilize \texttt{HEASOFT}. Utilizing imaging for the seeding would also likely limit the sensitivity of the search, particularly for bursts at low partial coding that rely on significant response from the uncoded detectors. However, it could be very helpful in reducing the number of positions to analyze inside each square position seed for sources that are not too weak.


\subsection{Likelihood Analysis}
\label{LikAnalysisSect}

With the seeds finally found, jobs to maximize the LLH (as outlined in Section \ref{OnTimeOpt}) for each position seed are distributed across the workers running on the computer cluster. For each square position seed, a grid of position points is made, as shown in Figure \ref{SplitRatesIMxyGRB180805B}. The grid spacing in IMY is 0.003 and in IMX is 0.004, where each row of IMX points is offset by 0.002. A non-square grid like this helps to minimize the distance between any position and the closest grid point. A 3x3 grid of spectral parameters is used with $\gamma\in$ [0.1, 0.6, and 1.1] and $E_{\rm peak}\in$ [97.7, 212.1, 460.6] keV. The worker finds the maximum LLH at each position point and for each time seed that have this square position seed. If any position is found to have a $\sqrt{\Lambda}>$ 6 a finer scan in position and spectral points are done around the top few max LLH `peaks' (positions separated by more than the PSF FWHM). For each time bin that passes the $\sqrt{\Lambda}$ cut, the peaks to scan around are found by calculating the $\Delta LLH$ from the maximum LLH position. Any positions more than 0.009 image units away from another peak and with $\Delta LLH<$ 10 are added to the list of peaks to scan, up to a maximum of 4 peaks and including at least 2 if there's nothing with $\Delta LLH<$ 10. For each peak the LLH is maximized for positions in a 3x3 square grid with steps of 0.002 in IMX and IMY around the peak position and a 3x3 grid of spectral points around the best fit spectral point for the peak. Then, around the peak with the largest LLH after those scans one more fine grid scan is done with steps of 0.001 in IMX and IMY and slightly smaller steps in spectral points. An example of this recursive grid scanning with GRB 180805B is shown in Figure \ref{SplitRatesIMxyGRB180805B}. Each worker does this for each square position seed assigned to it. 

For the OFOV part of the search, for each time seed the same analysis is performed for each sky position at partial coding fraction $<5$\%. This sky positions are determined via a HEALPIX map with an NSIDE of 16 (position spacing of $\sim4$ degrees), so there is some possible overlap between the in and out of FOV results. Unlike the IFOV search, there is no recursive position grid scanning, but the recursive finer scanning is still performed for the spectral grid points. For the real-time search, the manager process running on the AMON servers monitors the search processes and gathers and reports the results once all the workers have finished.

\subsection{Example Pipeline Results}

\begin{figure}[htb]
	\centering\includegraphics[width=6 in]{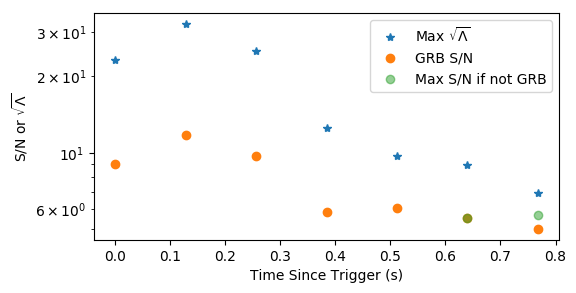}
	\caption{The $\sqrt{\Lambda}$ values for seven 0.256 s durations around GRB 180805B from the pipeline, as well as the maximum SNR values found by \texttt{batcelldetect} in the same seven times. The green markers indicate when the maximum SNR position from \texttt{batcelldetect} is \textit{not} at the true GRB position. The second to last SNR time bin had the GRB as the second highest SNR by 0.02. The maximum LLH was found at the true GRB position for each time bin.}
	\label{TS_SNR_tbins}
\end{figure}

In this section, we demonstrate the results of the pipeline run on GRB 180805B, with the only difference from the real-time pipeline being that specific time bins are chosen, instead of the time seeding in order to get results for more time bins for comparison to imaging. Every other piece of the pipeline is kept identical. The search was run on seven 0.256 s duration time bins with steps of 0.128 s. For each time bin the GRB's position is successfully picked in the position seeding process and subsequently found as the maximum likelihood position. The resulting $\sqrt{\Lambda}$ for each time is plotted in Figure \ref{TS_SNR_tbins}. The maximum $\sqrt{\Lambda}$ found was 32.37 in the second time bin. 

At the same time bins, the standard imaging analysis was performed to compare the results. The images were created with \texttt{batfftimage}, using a 20 s background detector plane image for energies 15 keV - 350 keV. Then, \texttt{batcelldetect} is used to search each image for points sources down to a SNR of 3.5. The true GRB position is found to have the highest SNR of any unknown source found in the first 5 images, but not in the last 2. For those time bins the GRB position is unrecoverable with conventional imaging. The maximum SNR values and the SNR nearest to the GRB are plotted in Figure \ref{TS_SNR_tbins}. 

The $\sqrt{\Lambda}$ from the likelihood search is substantially higher than the SNR at all times, showing how the likelihood analysis is more sensitive than imaging. These results also show that the likelihood analysis is able to recover the true GRB position even when conventional imaging would fail to discriminate the correct position.

\subsection{Interpreting the test statistic $\sqrt{\Lambda}$}
\label{InterpLambda}

We would like to be able to convert $\sqrt{\Lambda}$ to an interpretable significance, like a p-value or a false alarm rate (FAR). However, we are prevented from relying on an analytical statistical form. This is for several reasons; the tail end of the $\Lambda$ distribution is less likely to behave well, there are many overlapping trials, and the non-GRB outliers in the distribution are most likely not statistical fluctuations but instead caused by glitches that were not successfully screened, or fluctuations in the strength of other background sources or the local particle background. For this reason we must rely on analysis results using real data that does not include a BAT-detected GRB in order to generate a distribution of $\Lambda$. TTE data is usually only available when there is a burst detected onboard, or from GUANO also targeted specifically at possible burst times, but there also exists some TTE data that is kept for calibration. Using data from calibration runs, other pre-planned TTE data recordings, and data from before or after known GRB signal times, 51.2 ks of TTE data is assembled to pass through the analysis pipeline. There is more usable TTE data in the archive, likely  $\sim 250$ ks, but it takes a very long time to run the pipeline on this much data. 


\begin{figure}[htb]
	\centering\includegraphics[width=6in]{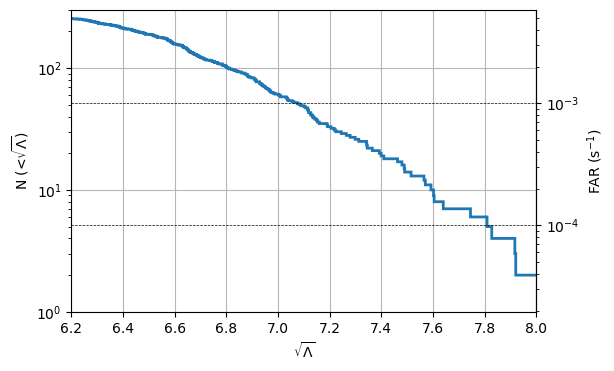}
	\caption{The reverse cumulative distribution of the resulting $\sqrt{\Lambda}$ values from the pipeline being run on 51.2 ks of TTE data. The right y-axis shows the the left y-axis dived by 51200 s to get a rate.}
	\label{FARplot}
\end{figure}

The pipeline is run on 40 s chunks of data, stepping over the total 51.2 ks, just like the regular search. Figure \ref{FARplot} shows the reverse cumulative distribution of the maximum $\sqrt{\Lambda}$ found in each 40 s search. The FAR as a function of $\sqrt{\Lambda}$ is calculated as the number of occurrences of a $\sqrt{\Lambda}$ greater or equal to that, divided by the total time analyzed. It should be noted that this data is not a perfect representation of randomly selected data. Times periods during which TTE data is selected to be kept for calibration, and when there was a GRB recently detected, are less likely to present problems like an unusual background or be close to the SAA. For this reason, this distribution may exclude those times that could cause anomalously high $\sqrt{\Lambda}$'s, but these issues are typically found on a manual inspection. 

Two outliers were found in the background $\sqrt{\Lambda}$ distribution. The largest outlier was at a $\sqrt{\Lambda}$ value of 15.07 and was caused by an OFOV GRB (\fermi\ trigger 443463980\footnote{\url{https://gcn.gsfc.nasa.gov/other/443463980.fermi}} and INTEGRAL trigger 6867\footnote{\url{https://gcn.gsfc.nasa.gov/other/6867.integral_spiacs}}). The other was a noise event at a $\sqrt{\Lambda}$ value of 9.32. Upon manual inspection the noise event was found to be caused by several high-energy cosmic ray events happening inside of a single 0.128 s time bin that were each too weak individually to be caught by the glitch detection in the data cleaning step. Noise events like this are caught upon manual inspection, but could be caught automatically with future improvements to the glitch screening. Based on this constructed population, a noise event occurring with a $\sqrt{\Lambda}>$  8 is rare and most likely occurs at a rate comparable to the true rate of detectable GRBs.

The position of the maximum LLH location helps us determine if the burst was inside or outside of the FOV. This is particularly useful because if we know the burst originated from outside the coded FOV then we know there's no chance of localizing it to an arcminute-scale for followup observations. If the maximum LLH value is found outside the FOV this typically means that the burst's true position is outside of the FOV. However, the case is less clear if the maximum LLH value is found inside the FOV. The same coded noise that creates noise peaks in images ranging from a SNR of $\sim -5\sigma$ to $\sim 5\sigma$ in a typical image also affects the likelihood inside the FOV. Due to the coded noise, and significantly more trials, weakly detected bursts originating from outside of the FOV tend to have their maximum LLH position found inside the FOV. The probability that a burst truly originates outside the FOV is assessed using the $\Delta LLH_{out} = max(LLH_{in}) - max(LLH_{out})$ statistic, where negative values mean the burst is most likely outside the FOV, large positive values mean the burst is most likely inside the FOV, and small positive values mean the burst's location is more ambiguous. Typical values of ``small'' and ``large'' positive $\Delta LLH_{out}$ can be seen in section \ref{SectDiscoveries}.

\section{Sensitivity of NITRATES and detection range for GRB 170817A}
\label{sec:sensitivity}
In order to determine the sensitivity of NITRATES, simulated events are injected into existing data and the search is run to see if the simulated GRB is recovered. The simulated events are generated by first finding the count expectation in each detector and energy bin from the response, given the simulated GRB's position, spectrum, flux, and duration. Then, the number of simulated events in each detector and energy bin are determined via a Poisson random draw given the count expectations. The event times are randomly assigned between some start time and the end of the signal duration. Finally, the simulated events are injected into an existing TTE data file and the search is run to find the maximum $\sqrt{\Lambda}$, with the spatial parameter space limited to  positions near the simulated GRB to save computation time. This is done many times, with small, random perturbations to the GRB's position and start time. The sensitivity for a given position and spectral parameters is determined by finding at what simulated flux does the search find the GRB at a $\sqrt{\Lambda} \geq$ some value, 50\% of the time. Here we use $\sqrt{\Lambda} \geq 7.5$, which corresponds to a FAR of $\sim$ 1 per hour. It should be noted that this is the detection sensitivity and not the sensitivity for acquiring an arcminute localization. Localizing an IFOV GRB to a single arcminute-scale peak may require a slightly higher fluence.

We used this method of determining the search sensitivity to find the maximum recoverable distance for the main peak of GRB 170817A. The GBM lightcurve binned in 0.128 s bins and the best-fit spectral parameters ($E_{\rm peak}$ = 185 keV and $\alpha$ = -0.62) were used to simulate GRB 170817AA at two positions in the BAT FOV. The data that the simulation was injected into had $\approx$18,000 detectors active at the time. At the center of the BAT FOV we found a maximum recoverable distance of $\sim$100 Mpc, and at $\theta = 45^{\circ}$, $\phi = 0^{\circ}$ we found a maximum recoverable distance of $\sim$86 Mpc. This is assuming a luminosity distance of 41 Mpc (\cite{grbDistCant}; \cite{grbDistHjorth}) to GRB 170187A.  It should be noted that including skymap information from a GW detection with a suitable joint-detection statistic can further increase the recoverable range, for a given choice of False-Alarm-Rate (FAR) threshold (Tohuvavohu, Ewing, et al. in prep). 

In contrast, GRB170817A would not trigger \swift/BAT onboard beyond $\sim65$ Mpc \citep{GUANO}, and would not be detected via convential imaging on the ground to beyond $\sim75$ Mpc in the best case. To provide context with respect to other GRB missions, GRB 170817A would not trigger \fermi/GBM onboard beyond 50 Mpc \citep{GBMgrb170817A}, and would not be detectable by the GBM targeted search ground analysis beyond $\sim74$ Mpc \citep{GBMsubthresh}. However, \fermi/GBM has significant sensitivity over the entire unocculted sky, so a direct comparison to IFOV BAT detections results in a higher accessible volume for \fermi/GBM of $\sim$ 3x, despite the restricted range. We have not assessed sensitivity and recovery distances for GRB 170817-like events originating from outside the BAT coded FOV, despite significant sensitivity across the unocculted sky (see Figure \ref{fig:AEffMap}), due to remaining large systematic uncertainties in these responses. For this reason, an apples-to-apples detectable volume comparison with \fermi/GBM is not currently feasible. Despite this, \swift/BAT uniquely provides arcminute localizations for bursts inside its coded FOV and therefore enables early time followup observations, unlocking physics that would not be otherwise accessible.

\section{GUANO NITRATES enabled discoveries}
\label{SectDiscoveries}

The first successful recovery of BAT TTE data by GUANO for a scientific trigger was in April 2019, triggered by GW190408\_181802 \citep{s190408gcn,gwtc2}. At this time, GUANO was still in development stages, and recovery efficiency remained low throughout O3a. As the commanding efficacy of GUANO and supporting infrastructure was improved, and moved to full automation, TTE recovery became more consistent for the second half of the run (O3b) (see discussion and Figure 5 in \citealt{GUANO}). Low-latency targeted searches with NITRATES were run on hundreds of above, and below, threshold GW triggers from the LVC during this period (Tohuvavohu et al. 2022, in prep). In February 2020 GUANO also began accepting triggers to dump data for alerts from other GRB detectors with the hope of localizing bursts for follow-up that were in the BAT FOV, but were too weak or otherwise missed by the onboard trigger algorithms, as well as building up the number of out of FOV bursts with TTE data for the purpose of calibrating the response. GUANO now responds to an average of 5 external triggers per day (GRBs, GWs, FRBs, neutrinos), achieving $\sim90\%$ TTE data recovery rate. All of this data is subsequently processed by NITRATES. The yield from external GRB triggered searches has been substantially higher than initially expected, yielding GUANO NITRATES derived arcminute GRB localizations at a rate of $\sim1$/month.

\subsection{NITRATES search outputs description}
The search pipeline returns results and helpful contextual information for each signal hypothesis tested and likelihood evaluation, for the various separate phases of the full targeted search. We remind our reader that for the purposes of evaluating the significance of a detection and localization, the most useful metrics are:
\begin{itemize}
    \item $\sqrt{\Lambda}$ (sqrt-TS), which is the likelihood ratio test statistic, reporting the evidence for the hypothesis of a GRB in the data, relative to the background-only hypothesis. \\
    \item $\Delta LLH_{out}$, defined as the difference in log-likelihood between the maximum LLH peak for an IFOV location vs an OFOV location. This value describes the degree of preference for an in vs out of FOV origin for the burst. \\ 
    \item $\Delta LLH_{\rm peak}$, defined as the difference in log-likelihood between the maximum LLH peak and the separate spatial peak that has the next highest LLH. This value describes the degree of preference for a particular location for the signal within the FOV. \\
\end{itemize}

\subsection{GRBs Detected by BAT-GUANO NITRATES}

\begin{figure}[htb]
	\centering\includegraphics[width=6 in]{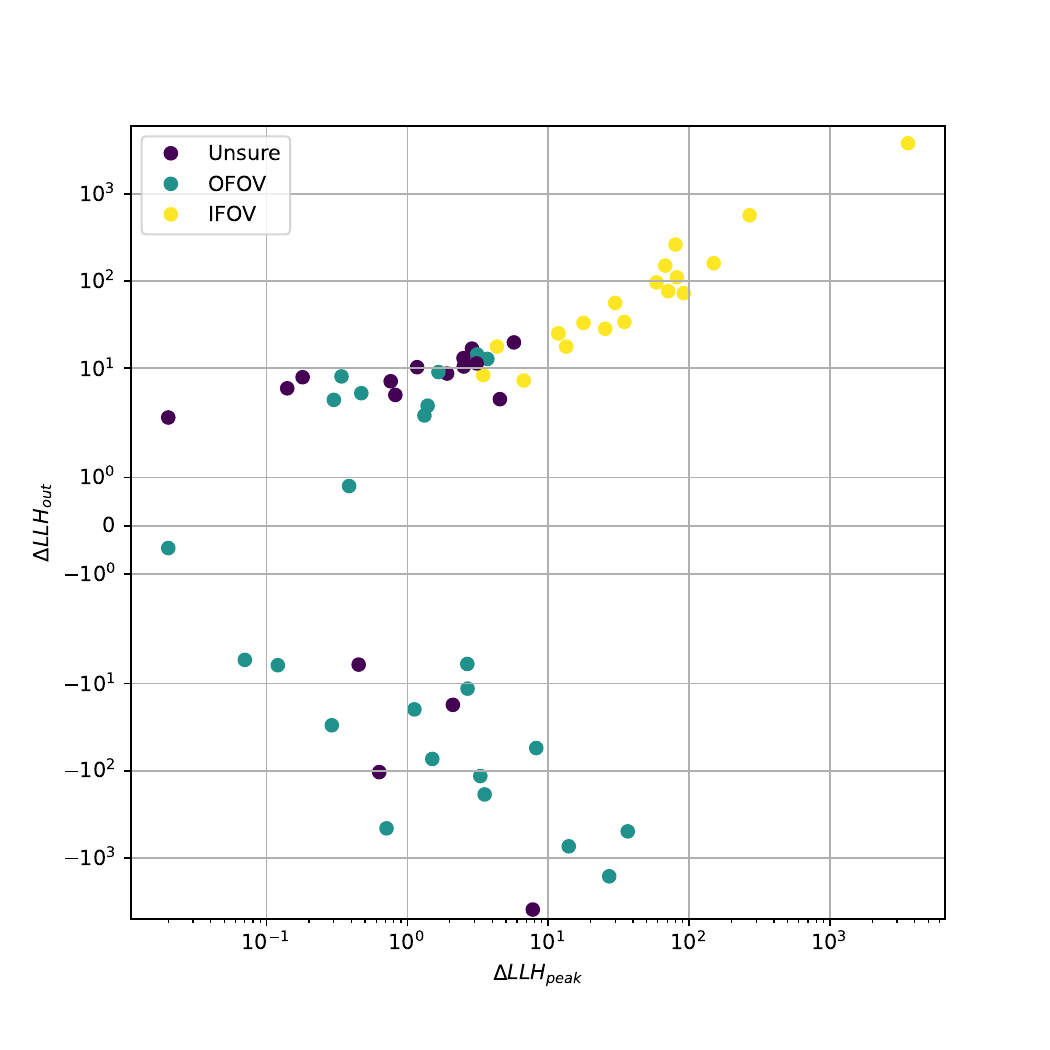}
	\caption{The $\Delta LLH_{out}$ and $\Delta LLH_{\rm peak}$ values for bursts from Table \ref{table:AllResults} and \ref{table:LocalizedResults} color coded by the In or Out FOV designation in Table \ref{table:AllResults} and marked `Not Sure' if it could not be ruled In or Out FOV. }
	\label{DloglOutVsPeakLog}
\end{figure}

The summary of the search results for GRBs found in the period from February 2020 to August 2021 is shown in Table \ref{table:AllResults}. The table also includes whether each burst can have the portion of sky covered by the BAT coded FOV ruled out from a localization made using other instrument(s) and if the burst can be localized to be contained inside the BAT coded FOV. There are 14 bursts without either of these marks that were not localized in this search and could be either inside or outside the FOV. In Figure \ref{DloglOutVsPeakLog} the results can be seen in the $\Delta LLH_{out}$ vs $\Delta LLH_{\rm peak}$ parameter space. Only GRBs marked as ``IFOV'' have positive $\Delta LLH_{out}$ and $\Delta LLH_{\rm peak}$ values $>$ 10, so we tentatively mark our threshold for confident arcminute scale localizations at a $\Delta LLH_{\rm peak}$ of 10. For detections with positive $\Delta LLH_{out}$ values and $\Delta LLH_{\rm peak}$ values $<$ 10 we find a mix of detections marked ``IFOV'' and ``Unsure'', along with some marked ``OFOV'' at $\Delta LLH_{\rm peak}$ values closer to zero. We report detections in this region of parameter space as candidate localizations. These candidate localizations can be promoted to a confident localization by either finding a spatially coincident, fading afterglow with follow up observations or a spatially coincident image peak that is at least marginally significant (SNR $\gtrsim$ 6) on its own.  

\begin{sidewaystable}\scriptsize
\begin{tabular}{cccccccccccccc}
Trigger Time & Name & T$_{90}$ & RA\footnote{\label{GBMcatCoords}\scriptsize Listed localization in GBM burst catalog} & DEC\textsuperscript{\ref{GBMcatCoords}} & Error\textsuperscript{\ref{GBMcatCoords}} & $\Delta\Psi_{pnt}$\footnote{\scriptsize The angular distance between the BAT pointing direction and the listed localization.} & $\sqrt{\Lambda_{in}}$ & $\sqrt{\Lambda_{out}}$ & $\Delta LLH_{\rm peak}$ & $\Delta LLH_{out}$ & Exp.\footnote{\scriptsize The exposure of the max LLH found in the search.} & OutFOV\footnote{\scriptsize Whether an external localization can rule that the burst was definitively outside BAT's coded FOV.}  & InFOV\footnote{\scriptsize Whether the localization from BAT or another detector can rule that the burst was inside the BAT's FOV.} \\
&  & s & J2000 $\mathrm{{}^{\circ}}$ & J2000 $\mathrm{{}^{\circ}}$ & $\mathrm{{}^{\circ}}$ & $\mathrm{{}^{\circ}}$ &  &  &  &  & s & &\\ \hline
2020-02-16 09:07:25 & GRB200216380 & 8.192 & 311.438 & -11.658 & 0.05 & 32.4 & 18.59 & 16.77 & 34.71 & 33.89 & 8.192 & & X \\
2020-02-28 06:58:33 & GRB200228291 & 3.584 & 328.01 & -46.45 & 1.51 & 32.0 & 161.89 & 136.40 & 3585.22 & 3801.66 & 2.048 & & X \\
2020-03-18 22:35:27 & GRB200318941 & 10.24 & 123.45 & 59.07 & 1.0 & 81.3 & 11.38 & 10.33 & 3.69 & 12.74 & 8.192 & X & \\
2020-03-25 03:18:31 & GRB200325138 & 0.96 & 31.7203 & -31.816 & 0.07 & 55.5 & 27.02 & 25.95 & 25.38 & 28.28 & 1.024 & & X \\
2020-03-27 20:56:48 & GRB200327873 & 0.64 & 236.54 & -4.217 & 3.6 & 77.5 & 21.45 & 21.21 & 0.47 & 5.15 & 0.256 & X &  \\
2020-04-05 03:53:38 & GRB 200405B & & & & & & 12.41 & 10.39 & 2.87 & 16.73 & 0.128 & & \\
2020-04-15 08:48:05 & GRB200415367 & 0.144 & 11.885 & -25.263 & 0.12 & 48.5 & 101.66 & 102.78 & 3.29 & -115.21 & 0.128 & X & \\
2020-05-07 15:52:58 & GRB200507662 & 55.297 & 57.83 & 65.52 & 2.2 & 61.3 & 13.78 & 13.44 & 0.82 & 4.92 & 0.512 & & \\
2020-05-30 00:44:10 & GRB200530031 & 22.528 & 32.1 & 69.23 & 0.49 & 106.1 & 15.17 & 14.98 & 1.32 & 2.86 & 16.384 & X & \\
2020-06-01 02:19:50 & GRB200601097 & 153.859 & 36.71 & 33.49 & 1.73 & 72.0 & 25.45 & 25.90 & 2.67 & -11.43 & 16.384 & X & \\
2020-06-05 18:17:42 & GRB200605762 & 0.32 & 95.839 & 50.954 & 0.6 & 142.8 & 18.54 & 18.34 & 1.39 & 3.70 & 0.128 & X & \\
2020-06-07 08:41:49 & GRB200607362 & 20.48 & 171.5 & 30.88 & 2.1 & 78.1 & 16.12 & 15.20 & 3.12 & 14.37 & 16.384 & X & \\
2020-06-07 22:06:31 & GRB200607921 & 26.625 & 41.33 & -83.33 & 1.0 & 125.6 & 23.14 & 23.40 & 2.66 & -5.96 & 4.096 & X & \\
2020-06-13 05:30:08 & GRB200613229 & 478.026 & 153.042 & 45.7541 & 0.0004 & 126.5 & 31.56 & 33.80 & 1.50 & -73.34 & 4.096 & X & \\
2020-06-19 02:36:10 & GRB200619108 & 28.672 & 101.35 & 56.74 & 4.02 & 76.1 & 32.75 & 32.93 & 0.45 & -6.06 & 8.192 & &  \\
2020-06-22 02:11:48 & GRB200622092 & 32.001 & 107.52 & 38.28 & 1.35 & 72.6 & 27.91 & 28.61 & 1.12 & -19.78 & 16.384 & X & \\
2020-06-23 03:18:00 & GRB200623138 & 0.096 & 242.095 & 53.4678 & 0.07 & 23.9 & 10.09 & 9.23 & 3.45 & 8.30 & 0.128 & & X \\
2020-07-14 08:15:25 & GRB200714344 & 0.96 & 190.43 & -53.95 & 2.17 & 76.0 & 12.54 & 12.06 & 0.14 & 5.86 & 0.256 & & \\
2020-07-16 01:26:37 & GRB200716060 & 45.057 & 134.83 & -15.02 & 1.0 & 55.1 & 110.96 & 110.1 & 81.86 & 110.10 & 16.384 & & X \\
2020-08-03 16:52:43 & GRB200803703 & 54.784 & 198.5 & -42.5 & 4.7 & 61.6 & 19.22 & 18.99 & 4.53 & 4.40 & 16.384 & & \\
2020-08-09 15:41:27 & GRB200809654 & 17.408 & 15.8295 & -73.8456 & 0.0013 & 43.4 & 39.12 & 37.23 & 91.60 & 72.35 & 1.024 & & X \\
2020-09-07 22:57:28 & GRB200907957 & 140.803 & 34.5 & -78.38 & 1.15 & 143.5 & 40.63 & 41.97 & 8.21 & -54.97 & 8.192 & X &  \\
2020-09-23 17:57:43 & GRB200923748 & 0.64 & 126.3 & -54.25 & 11.9 & 55.9 & 10.44 & 9.40 & 1.17 & 10.24 & 0.128 & &  \\
2020-10-08 10:37:36 & GRB201008443 & 2.048 & 160.48 & 47.65 & 3.33 & 46.9 & 12.90 & 12.47 & 4.32 & 17.65 & 4.096 & & X \\
2020-10-16 00:27:53 & GRB201016019 & 2.944 & 174.3 & -1.96 & 1.0 & 89.1 & 115.54 & 128.87 & 27.08 & -1628.78 & 2.048 & X & \\
2020-10-20 17:33:54 & GRB201020732 & 15.872 & 71.55 & 76.55 & 1.62 & 63.0 & 136.12 & 141.44 & 13.97 & -737.51 & 4.096 & X & \\
2020-11-16 00:49:47 & GRB201116035 & 6.784 & 149.48 & -4.2 & 1.0 & 69.3 & 105.54 & 110.15 & 36.67 & -497.07 & 8.192 & X & \\
2020-12-16 19:00:31 & GRB201216792 & 77.312 & 201.49 & 36.31 & 0.05 & 41.9 & 22.88 & 19.4 & 80.06 & 261.66 & 16.384 & & X \\
2020-12-18 04:14:15 & GRB201218177 & 14.336 & 286.88 & 24.18 & 1.0 & 120.9 & 85.01 & 90.24 & 0.71 & -458.90 & 8.192 & X & \\
2020-12-27 15:14:07 & GRB201227635 & 0.096 & 170.121 & -73.613 & 1.0 & 61.2 & 66.44 & 66.89 & 0.29 & -30.10 & 0.128 & X & \\
2020-12-28 15:21:45 & GRB201228640 & 37.121 & 8.14 & 50.69 & 8.56 & 50.8 & 27.88 & 21.59 & 149.51 & 159.92 & 16.384 & & X \\
2021-05-06 00:39:48 & I9191 & - & - & - & - & - & 18.03 & 17.98 & 0.385 & 0.822 & 0.128 & X &  \\
2021-01-23 07:19:03 & GRB210123305 & 13.056 & 345.12 & -56.93 & 1.14 & 136.6 & 21.81 & 22.09 & 0.12 & -6.16 & 4.096 & X &  \\
2021-02-27 02:45:09 & GRB210227115 & 17.92 & 115.76 & 61.11 & 5.8 & 62.2 & 12.98 & 12.77 & 0.02 & 2.71 & 8.192 &  &  \\
2021-03-03 10:11:38 & GRB210303425 & 13.056 & 44.53 & -3.78 & 5.24 & 63.5 & 10.56 & 9.79 & 0.18 & 7.86 & 8.192 &  &  \\
2021-03-14 11:13:45 & INTEGRAL 9081 &  &  &  &  &  & 10.27 & 9.55 & 0.76 & 7.06 & 0.128 &  &  \\
2021-03-19 01:55:38 & CALET 1300154125 &  &  &  &  &  & 12.26 & 11.54 & 1.91 & 8.66 & 2.048 &  &  \\
2021-03-23 12:03:33 & GRB210323502 & 13.568 & 259.687 & 15.634 & 0.0013 & 54.5 & 11.75 & 10.83 & 2.51 & 10.42 & 8.192 &  &  \\
2021-05-04 23:00:47 & GRB210504959 & 13.568 & 65.06 & 49.06 & 3.91 & 64.7 & 25.62 & 26.29 & 2.10 & -17.55 & 8.192 &  &  \\
2021-05-24 05:00:04 & GRB210524208 & 20.992 & 337.32 & -47.05 & 1.53 & 110.8 & 10.83 & 9.96 & 1.66 & 9.00 & 8.192 & X &  \\
2021-06-02 12:04:01 & GRB 210602A &  &  &  &  &  & 17.29 & 17.59 & 0.07 & -5.35 & 1.024 & X &  \\
2021-06-05 05:08:58 & GRB210605215 & 0.448 & 21.67 & -41.51 & 2.01 & 79.7 & 24.16 & 24.18 & 0.02 & -0.46 & 0.256 & X &  \\
2021-06-06 03:56:02 & GRB210606164 & 11.008 & 170.904 & 0.718 & 0.066 & 42.8 & 14.90 & 12.50 & 17.78 & 32.97 & 8.192 &  & X \\
2021-06-06 22:41:07 & GRB210606945 & 243.717 & 87.48 & -18.41 & 1.2 & 71.2 & 91.12 & 93.16 & 3.53 & -187.24 & 2.048 & X &  \\
2021-06-07 21:39:15 & GRB 210607C &  &  &  &  &  & 244.72 & 260.24 & 7.76 & -3919.97 & 16.384 &  &  \\
2021-06-22 01:32:35 & GRB210622064 & 40.704 & 233.117 & -26.213 & 0.0417 & 38.5 & 18.89 & 17.93 & 13.41 & 17.66 & 2.048 &  & X \\
2021-06-27 17:57:20 & GRB210627748 & 15.68 & 234.84 & -5.97 & 2.13 & 96.6 & 10.41 & 9.61 & 0.34 & 8.02 & 1.024 & X &  \\
2021-07-03 07:16:25 & CALET 1309331577 &  &  &  &  &  & 31.57 & 34.70 & 0.63 & -103.76 & 8.192 &  &  \\
2021-08-02 20:08:06 & GRB210802839 & 10.24 & 229.92 & 30.3 & 1.0 & 140.9 & 21.13 & 20.92 & 0.30 & 4.32 & 8.192 & X &  \\
2021-08-27 09:59:24 & GRB210827416 & 8.704 & 174.918 & 55.786 & 0.066 & 29.4 & 17.34 & 13.75 & 29.86 & 55.88 & 8.192 & & X \\
\end{tabular}
\caption{Summary of the GRB search results through August 2021.}
\label{table:AllResults}
\end{sidewaystable}

\subsection{Localized GRBs}
\label{LocalizedGRBs}

Of the $>100$ GRBs detected in NITRATES analyses run on GUANO data, arcminute localizations have been determined for 32 bursts as of July 2022. Of these, 23 were found by the NITRATES pipeline described here. The other 9 occurred during periods of spacecraft slew, where the attitude is not stable, and the current version of NITRATES is not capable of analysing these. Those 9 bursts were found with mosaicked conventional imaging a la \cite{BATSS}. The arcminute localizations were all announced in GCN Circulars with latency ranging from 3 hours to 2 days post burst, with a median latency of 11 hours, set by the current compute resources available for the search and other activity on the cluster. Many of our reported localizations received followup by other instruments for afterglow searches, of which 13 have been found. The results are summarized in Table \ref{table:LocalizedResults}, where the significance of the NITRATES detection and localization can be compared to that found from conventional imaging.

In the cases where NITRATES produced a candidate arcminute-scale localization, standard imaging was also attempted to try to 1) confirm the location and 2) provide a more accurate location. The likelihood analysis uses responses from linear interpolations of forward ray traces that are 0.002 image units apart ($\approx$ 0.1$^\circ$), which limits the accuracy of derived localizations. However, \texttt{batcelldetect} uses a proven method of fitting the PSF to the image. For this reason, even at low SNR where it may not be possible to determine confident location from imaging alone, it is preferred to use the associated \texttt{batcelldetect} source location, if available. In these cases the NITRATES analysis is effectively discriminating between low SNR sources in the image, even when the image source associated with the true GRB location does not have the highest SNR. In some cases no nearby source is found with imaging even down to very low SNR (see eg GRB 210323B), and in these cases we rely on the NITRATES location only. When attempting to find the source location using \texttt{batcelldetect}, we do a blind search for sources down to a SNR of 3.5, which results in $\approx$ 200 - 250 source candidates that are all most likely from noise. If any source candidates are found within 0.2$^\circ$ of the NITRATES position, we report the source candidate's position and error (found as a function of SNR\footnote{\url{https://swift.gsfc.nasa.gov/analysis/bat_digest.html}}) as the GRB's best localization. Even though there are many candidates found from noise using \texttt{batcelldetect} at this low threshold, there is only a $\approx$ 0.5\% random chance of finding a noise candidate within 0.2$^\circ$ of the NITRATES position per image searched.

NITRATES was run fully autonomously, using the pre-specified parameters. However, for imaging analysis several different exposures and energy bins were tried until the SNR was maximized to have the best localization. In the following subsections we provide brief descriptions of a few selected bursts that either exemplify the sensitivity and discriminatory power of NITRATES, or are good examples of marginal events for whom localizations remain uncertain. 
\begin{table}[h]\footnotesize
	\caption{Arcmin Localized GRB Results}
	\hspace*{-3cm}
	\begin{tabular}{cccccccccc}
		NAME & RA & DEC & SNR & $\sqrt{\Lambda}$ &  $\Delta LLH_{\rm peak}$ & Loc. Found & Followup Began & Afterglow? & Comments \\
		\hline
		GRB 200216A & 311.438 & -11.658 & 7.5 & 18.59 & 34.71 & T0+12 hr & Sun Constraint & NA &  \\
		GRB 200228A & 333.893 & -42.944 & 40 & 161.9 & 3585.2 & T0+8 hr & Sun Constraint & NA &  Missed onboard b/c late trig. restart\\
		GRB 200325A & 31.720 & -31.816 & 7.5 & 27.02 & 25.38 & T0+20 hr & Sun Constraint & NA & Agrees with small IPN loc. \footnote{\cite{GRB200325AgcnIPN}} \\
		GRB 200405B & 63.317 & -48.823 & 3.5 & 12.41 & 2.87 & T0+17 hr & T0+18.5 hr & N & Agrees with large IPN loc. \footnote{\cite{200405b-ipn}} \\
		GRB 200623A & 242.095 & 53.468 & 6.4 & 10.09 & 3.45 & T0+10 hr & T0+10.5 hr & N & Coming out of SAA \\
		GRB 200714E & 196.865 & -51.640 & 11.5 & -  & - & T0+9 hr & T0+10.5 hr & \textbf{Y} & During slew. XRT afterglow. \\
		GRB 200716A & 139.348 & -16.712 & 8.5 & 110.96 & 81.86 & T0+ 11 months & NA & NA & Agrees with small IPN loc \footnote{\cite{GRB200716AgcnIPN}} \\
		GRB 200809B & 15.941 & -73.846 & 10.4 & 39.12 & 91.60 & T0+3 hr & T0+3.5 hr & \textbf{Y} & XRT afterglow  \\
		GRB 201008A & 161.744 & 46.101 & 4.2 & 12.90 & 4.32 & T0+12 hr & T0+13 hr &  \textbf{Y} &  XRT afterglow\\
		GRB 201128B & 339.35 & -49.246 & 15 & - & - & T0+2 days & T0+2.8 days & N & During slew. \\
		GRB 201216A & 201.487 & 36.312 & 8.9 & 22.88 & 80.06 & T0+3 hr & T0+6 hr & N & \\
		GRB 201228B & 35.594 & 56.015 & 13.1 & 27.88 & 149.51 & T0+5 months & NA & NA & Found late \\
		GRB 210323B & 259.664 & 15.677 & - & 11.75 & 2.51 & T0+14 hr & T0+28 hr & N & No image source!\\
		GRB 210421B & 270.817 & 56.828 & 9.7 & - & - & T0+12 hr & T0+12.5 hr & N & During slew.\\
		GRB 210605B & 15.732 & 6.467 & 8.4 & - & - & T0+8 hr & T0+20 hr & \textbf{Y} & During slew. XRT afterglow.\\
		GRB 210606A & 170.904 & 0.718 & 5.2 & 14.9 & 17.78 & T0+13.5 hr & T0+17 hr & \textbf{Y} & XRT afterglow. \\
		GRB 210622A & 233.117 & -26.213 & 8.1 & 18.8 & 12.6 & T0+12 hr & T0+2 days & N & \\
		GRB 210626A & 221.6199 & -1.1512 & 7.6 & - & - & T0+7 hr & T0+8 hr & \textbf{Y} & During slew. XRT afterglow. \\
		GRB 210706A & 312.0124	& 13.3079 & 19.1 & - & - & T0+10 hr & T0+12 hr & \textbf{Y} & During Slew. XRT afterglow. \\
		GRB 210827A & 174.9181 & 55.7858 & 8.3 & 17.3 & 29.8 & T0+11 hr & T0+12 hr & \textbf{Y} & XRT afterglow.\\
		GRB 211024A & 28.053 & -6.973 & - & 10.4 & 2.5 & T0+2 days & T0+2.5 days & N & Short. No image source!\\
	    GRB 211106A & 343.6522 & -53.23 & 3.58 & 19.1 & 6.7 & T0+8 hr & T0+11 hr & \textbf{Y} & Short, XRT, Chandra afterglow!\\
	    GRB 211229A & 295.044 & 23.133 & - & 50 & 269 & T0+7.5 hr & T0+12 hr & N & \\
	    GRB 220107A & 169.761 & 34.139 & 9.3 & - & - & T0+5 hr & T0+7.5 hr & \textbf{Y} & During slew. Afterglow, redshift!\\
	    GRB 220117C & 286.457 & 16.791 & 6.1 & 19.5 & 58.7 & T0+2 days & NA & NA & No followup, \swift\ in Safe. \\
	    GRB 220210A & 345.778 & 61.826 & 11.4 & 30.2 & 71 & T0+4 hours & NA & NA & Agrees with LAT pos. No followup.\\
	    GRB 220310C & 290.069 & 40.253 & - & 9.7 & 5.7 & T0+21 hours & T0+29 hours & No & No image source!\\
	    GRB 220418B & 224.344 & -17.539 & 5.6 & 10.4 & 11.8 & T0+12 hours & T0+32 hours & No & Short.\\
	    GRB 220511A & 287.446 & 17.739 & 9.9 & - & - & T0+8.5 hours & T0+10 hours & \textbf{Y} & During slew. XRT afterglow.\\
	    GRB 220606B & 28.474 & -54.741 & - & 10.7 & 3.1 & T0+11 hours & T0+12.5 hours & N & Short. No image source! Marginal loc.\\
	    GRB 220708B & 248.606 & 36.3225 & 5.9 & 41.4 & 67.7 & T0+7 hours & T0+10 hours & \textbf{Y} & XRT afterglow.\\
	    GRB 220710A & 221.592 & 20.657 & 20 & - & - & T0+2.5 days & T0+3 days & \textbf{Y} & During slew. XRT afterglow.\\	    
		\hline
		GRB 210506A & 132.853 & 4.582 & 6.8 & 18.03 & 0.385 & T0+6 hr & T0+8 hr & N & \textbf{Retracted}\footnote{Listed position is for the max SNR found in the image. NITRATES preferred an out of FOV origin \citep{GRB210506A}} . \\
\end{tabular}
\label{table:LocalizedResults}
\end{table}

\subsubsection{GRB 200405B}
GRB 200405B was a short GRB detected by INTEGRAL SPI-ACS\footnote{\url{https://gcn.gsfc.nasa.gov/other/8579.integral_spiacs}} that triggered GUANO. The burst was detected in the BAT TTE data with a duration of $\sim$0.5 s, and was reported in \citep{GRB200405Bgcn} along with a ``possible'' NITRATES localization. This was from an early version of NITRATES that did not support analyzing positions outside of the FOV and had a much less robust response, especially at positions close to the edge of the FOV. Using the most recent version of NITRATES the search was redone resulting in a different max LLH position. The $\Delta LLH_{\rm peak}$ is only 2.87 and the imaging SNR is only 3.5, so this new position is still not a confident localization. The $\Delta LLH_{out}$ is 16.73, which is in the range of uncertain if the burst originates from inside or outside the FOV. Both the original and updated max LLH positions agree with the IPN localization, but it is a very large region of $\sim1600~\mathrm{deg}^2$ \citep{200405b-ipn}. An external spectral measurement of the burst could help rule out some of the possible localization region, but no such measurement has been reported. Given the low $\Delta LLH_{\rm peak}$ we can not report a confident arcminute localization for this burst. 

\subsubsection{GRB 210506A}
\label{sect:210506a}
GRB 210506A was detected by INTEGRAL SPI-ACS \footnote{\url{https://gcn.gsfc.nasa.gov/other/9191.integral_spiacs}}, and triggered GUANO in near real-time. The burst was detected as a short, hard, GRB with a duration of 0.2 s in the BAT TTE data from GUANO. A rapid conventional imaging analysis revealed an uncatalogued point source in the BAT FOV with a SNR of 6.8. This significance put it right on the edge of the BAT imaging threshold, and we reported the candidate arcminute location at T0+6 hours, with the caveat that the candidate position of the burst coupled with its observed spectrum in BAT would imply an anomalously hard intrinsic spectrum of the GRB \citep{GRB210506A}. At this time, the NITRATES results were not yet available as the analysis was still running on the cluster. As the NITRATES search completed, it became clear that the analysis preferred an OFOV origin for the burst, and did not find any preference for the IFOV position found in the imaging. We reported this and retracted the candidate arcminute localization \citep{GRB210506Aretract}. An IPN timing localization of $\sim160~\mathrm{deg}^2$, reported at T0+3 days, confirmed the OFOV origin of the signal. This was a compelling, real-time, demonstration of one of the powers of NITRATES, to successfully differentiate spurious near-threshold localizations from conventional imaging that would be otherwise misleading.

\subsubsection{GRB 210323B}
GRB 210323B was a long GRB detected by \fermi\ GBM\footnote{\url{https://gcn.gsfc.nasa.gov/other/638193818.fermi}} which triggered GUANO. The burst was weakly but confidently detected in the BAT TTE data with a $\sqrt{\Lambda}$ of 11.75 found by NITRATES. The max LLH position agreed well with GBM's large localization, but the max LLH position was only marginally favored over other positions with a $\Delta LLH_{\rm peak}$ of only 2.51. With a $\Delta LLH_{out}$ of 10.42 it was in the uncertain region of whether it originated from inside or outside the FOV. After several iterations of the imaging analysis at different durations and energy ranges, \texttt{batcelldetect} was unable to recover an image peak near the max LLH position. Upon manual inspection of the images there was a positive SNR at the max LLH position, but it was $\textless 3\sigma$ and did not have a clear peak like point sources detected at higher SNR regularly do. Despite no image source and the low $\Delta LLH_{\rm peak}$ a \swift\ ToO was requested since the position agreed with the GBM localization. The followup observations began later than usual (T0+28 hr) due to NITRATES having a long run-time and unfortunate schedule timing. No afterglow was recovered in the observation \citep{GRB210323BxrtGCN}, which can be due to either an incorrect localization or the X-ray afterglow having already faded below detection level. The \fermi\ GBM team has since reported their spectral fit for GRB 210323B \citep{GBMgrb210323B}, but with rather large errors. The best fit spectra at the max LLH position agrees with GBM's spectral fit and fluence, but the next best fit peaks and some of the out of FOV region also agree within GBM's spectral fit confidence limits. Given all of this we can say the max LLH position is the most likely arcminute-scale region for GRB 210323B's true position, but we can not confidently say that it is definitely within that region. The localization probability is spread across multiple peaks inside the FOV and possibly outside of the FOV too. Being able to quantify the localization probability across multiple peaks and to outside the FOV is still a work in progress. 

\subsubsection{GRB 201008A}
GRB 201008A was detected by \fermi/GBM \citep{GRB201008Afermi} with a duration of $\sim 2$ seconds, on the border of the short-long classification. At a very low partial coding of 3.5\%, the burst was weakly but confidently detected in the BAT TTE data from GUANO with a $\sqrt{\Lambda}$ of 12.90 found by NITRATES.
NITRATES found a location for the burst with a $\Delta LLH_{\rm peak}$ of 4.31, which is marginally significant. Imaging yielded a source in the same location, with SNR 4.2. This location was consistent with the \fermi/GBM localization, and the best fit spectral shape and fluence from NITRATES agreed extremely well with that from \fermi/GBM (GBM Team, private communication), with a best fit 10-1000~keV fluence of $1.03\times10^{-6}$ erg/cm$^2$, $E_{\rm peak}$ of 212.1 keV, and $\gamma$ of 0.9.

Until the spectral agreement was established, the NITRATES localization was considered too marginal. Because of this, our reporting \citep{GRB201008Agcn} and ToO were a bit late and XRT was not able to start observing until 46.9~ks post-burst. One unknown X-ray source was found and then with later observations found to have faded away \citep{GRB201008AgcnXRT}, confirming that this was the GRB's correct position and an afterglow discovery. This burst has the lowest imaging SNR and $\Delta LLH_{\rm peak}$ of any NITRATES localized burst with a confirmed afterglow thus far, and shows how NITRATES is capable of not only detecting but also localizing bursts that are too weak to find through imaging. 

\subsubsection{GRB 210606A}
GRB 210606A was a long GRB detected by \fermi\ GBM\footnote{\url{https://gcn.gsfc.nasa.gov/other/644644567.fermi}} and triggered GUANO. NITRATES significantly detected and localized the burst with a $\sqrt{\Lambda}$ of 14.9 and a $\Delta LLH_{\rm peak}$ of 17.78. XRT began followup observations at T0+17 hr and was able to easily identify the fading afterglow \citep{GRB210606Axrt}. Using the imaging analysis a SNR of 5.2 was found at the GRB's position. This burst is another example of NITRATES detecting and localizing a burst that was too weak to be found through imaging. 

\subsubsection{GRB 211106A}
GRB 211106A was a short GRB detected by INTEGRAL/SPI-ACS, which triggered GUANO. NITRATES detected the burst with a $\sqrt{\Lambda}$ of 19.1, and a $\Delta LLH_{\rm peak}$ of 6.7. This was a significant detection, but the arcminute localization was only marginally confident due to the $\Delta LLH_{\rm peak} < 10$. XRT followup found a fading X-ray afterglow at the NITRATES position, confirming this position as correct. The image SNR at this position was 3.58. This burst turned out to be quite interesting, with the first discovered mm-afterglow to a short GRB, and one of the most energetic and wide-angle short GRB jets detected to date \citep{211106A}. This burst is one of the most excellent examples of NITRATES detecting and localizing a burst that was too weak to be found through imaging, and confirming a localization that was sub-threshold even for NITRATES.

\subsection{Bright Out of Field of View GRBs}
\label{OutFOVres}

Typically coded-aperture instruments cannot localize bursts outside of the coded FOV. However, with calibrated OFOV responses the maximum likelihood OFOV position can be found with simultaenous localization-spectral fitting similar to BALROG \citep{BALROG}. Since the response does not vary on arcminute scale outside of the FOV, the localizations will typically be limited to $\mathcal{O}$(10s of degrees) on the sky, even with perfectly calibrated responses. Even then, it still needs to be at a location where the response for this analysis does not have any serious errors. BAT detects more GRBs from OFOV then in, but only some are sufficiently bright to yield useful OFOV localizations. Figures \ref{BrightOFOVmap}, \ref{GRB201020Bhpmap}, and \ref{GRB201016Ahpmap} show the $\Delta LLH$ from the maximum LLH at each skymap pixel in the out of FOV portion of the search for three of the most strongly detected out of FOV GRBs. With an exactly correct response and errors these skymaps could use Wilks' Theorem to draw localization contours. Instead, we would have to rely on finding the distribution of $\Delta LLH$'s to the true position of GRBs using a large sample of bursts that are bright enough to be seen outside the FOV. As the OFOV responses become better calibrated by more OFOV burst detections the $\Delta LLH$ distribution will become more well behaved.

\subsubsection{GRB 201020B}

\begin{figure}[htb]
	\centering\includegraphics[width=5.5 in]{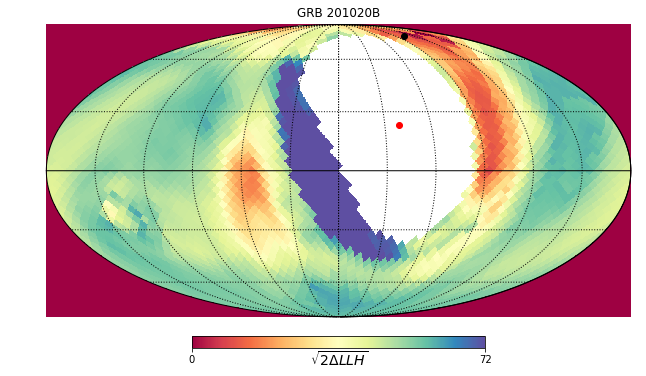}
	\caption{A HEALPix map with an Nside of 16 of the $\Delta$ LLH from the maximum LLH outside the FOV for GRB 201020B. The actual plotted value is $\sqrt{2\Delta LLH}$ for visualization purposes. The coded FOV is the blank portion of the map with the red dot showing the FOV center, and the black dot shows the GRB's position found by other detectors. The region of low $\Delta LLH$ centered around the GRB's position shows that the analysis did a good job of finding the GRB's position.}
	\label{GRB201020Bhpmap}
\end{figure}

GRB 201020B was found to be just outside the coded FOV ($\theta$ = 62.27$^\circ$, $\phi$ = 145.62$^\circ$), with an arcsecond-scale localization by XRT and several optical observations. It is clearly a long GRB, with several peaks that seem to have different spectral properties according to the Konus-Wind observations \cite{GRB201020BkwGCN}. The maximum LLH pixel in Figure \ref{GRB201020Bhpmap} isn't the closest pixel to the GRB's position, but it is pretty close and is contained within a strip of the largest LLH pixels. The spectral fit is hard to compare because of the changing spectral properties through the emission and the spectral fits by GBM and Konus-Wind were done for a Band spectrum, but it generally agrees. The max LLH point had a spectra with $E_{\rm peak}$ = 158.5~keV and $\gamma$ = 0.2, while the GBM fit to the whole emission \cite{GRB201020BgbmGCN} had an $E_{\rm peak}$ = 136.9~keV and a low-energy index of 0.71. 

\subsubsection{GRB 201016A}

\begin{figure}[htb]
	\centering\includegraphics[width=5.5 in]{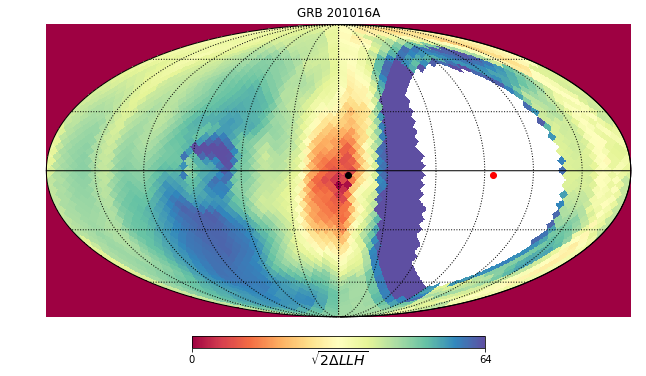}
	\caption{This figure is the same type of plot as Figure \ref{GRB201020Bhpmap}, but for GRB 201016A.}
	\label{GRB201016Ahpmap}
\end{figure}

GRB 201016A is a very bright, long GRB with a peak 1.024~s photon flux of 267.4 photons cm$^{-2}$ s$^{-1}$ (10 - 1000~keV) measured by GBM \cite{GRB201016AgbmGCN}. The GBM best-fit position puts the burst at $\theta =89.07^\circ$, $\phi = 261.26^\circ$ for BAT, which is one of the most complicated positions for the response as it goes through the XRT and UVOT tubes and is almost perpendicular to the detector array. The search ended up performing very well though
with the maximum LLH pixel just a few degrees away, which about what the GBM error is including systematics. The whole region in Figure \ref{GRB201016Ahpmap} with the lowest $\Delta LLH$ is around the GBM localization. The spectral parameters of the maximum LLH point seem a bit off with an $E_{\rm peak}$ = 63.1 keV and $\gamma$ = 1.8, but that might be from the spectral grid points being too coarsely spaced because another spectral point closer to what is expected from GBM's Band spectrum fit with $E_{\rm peak}$ = 181~keV and low-energy index of 0.42 had a just slightly smaller LLH value. 

\subsubsection{GRB 201116A}

GRB 201116A is another bright, long GRB with an arcsecond-scale localization from XRT and optical observations, placing it at $\theta = 65.76^\circ$, $\phi = 8.92^\circ$ for BAT. The analysis performed well again, with the maximum LLH pixel just a few degrees away from the GRB's position. In Figure \ref{BrightOFOVmap} (this was the example bright out of FOV burst from section \ref{sec:GRBsearch}), the whole high LLH region is again right around the GRB's position. The spectral parameters of the maximum LLH point were $E_{\rm peak} = 251.2$~keV and $\gamma = 1.4$, which is somewhat close to the Konus-Wind \citep{GRB201020BkwGCN} Band spectral fit with $E_{\rm peak}$ = 220~keV and low-energy index of 0.53. The offset in index may be from the coarse grid spacing again. 

\section{Improvements to NITRATES}
\label{sec:ImproveSearch}

\subsection{Calibration}

One of the goals of the GUANO data dumps for GRBs detected by other detectors is to obtain a large sample of TTE data for calibrating the OFOV response of BAT. The response for the likelihood analysis is kept in separate parts for different effects, detector groups, and source positions (see Appendix), which makes it easy to apply any correction factors or different errors for any of those parts of the response. As data for more out of FOV bright bursts with well fit spectra by other detectors become available, how the simulated response differs from the observed data can be explored to try to find which parts need corrections. Bright GRBs that were inside the FOV but at lower partial coding fractions can also be used to help examine the uncoded response, and especially at the boundary between uncoded and coded detectors where there are known issues, such as gaps between the mask and shield and parts of the mask that are blocked by other objects. Improved response calibration would lead to the $\Delta LLH$ as a function of sky position becoming more well behaved, which could allow for systematic (large) localization of GRBs outside of the coded FOV, as well as improving the overall sensitivity of the search. 

\subsection{Using Priors}

When performing the search around a GRB detected by another instrument (eg. \fermi/GBM) there can often be spectral fits available. Priors for the spectral norm and shape parameters from the best fit spectrum could be included in the NITRATES likelihood search. It would be important to make sure the priors are not too restrictive as reported errors often do not include systematics, and it would also be important to either match the time window of the other instrument or take into account more uncertainty from possible spectral evolution during the emission, with a much larger uncertainty for the spectral norm. To get $\Lambda$ the likelihood for the signal plus background model would need to be integrated over the spectral priors, which for a single position can be done by first maximizing the likelihood times the priors and then integrating in the region around the maximum point. The integrated likelihood should behave similarly when finding the maximum likelihood position and can be used inside $\Lambda$ with the background-only likelihood. Using these priors should lead to a more sensitive search, and a much better check than just seeing if the maximum likelihood position and spectral parameters generally agree with those from the other detector, as is current practice. 

Spectral parameter priors could also be used in targeted searches to improve sensitivity certain types of bursts. For example when looking for short-hard GRBs the distribution of $E_{\rm peak}$ and $\gamma$ values for bursts like that could be used to make the prior. Another example might be a search looking for soft gamma-ray repeaters (SGRs) using a thermal spectrum with a black-body energy prior taken from a distribution made from previously observed parameters for SGRs. 

\subsection{Using Other TTE Files}

Besides GUANO data dumps, limited amounts of TTE data are also saved when there is a failed onboard rate-trigger (no point source was found) and has at times been saved during some slewing periods. The TTE data from failed rate-triggers are short, containing only 3 or 10 s of data, which limits their utility and is too short for NITRATES in its current configuration due to insufficient time for background modeling. The slew data can be used for slew image mosaicking, but the changing attitude also makes it unsuitable for NITRATES as currently configured. 

Extending NITRATES to be capable of analyzing these short TTE data segments from `failed' rate triggers would enable recovery of GRBs in \swift's data archive, from before GUANO came online. However, using this data would require a way to fit a background model outside the burst time with at most 10 s of data. A background fit could be tried using the very limited non-signal time available, but fit errors would dominate. A better approach would be to simultaneously fit the background and signal parameters by having two time bins in the likelihood, a signal time bin where the background plus signal model will be and a background-only time bin which will consist of the remaining time outside the signal bin. It would help to reduce the number of energy bins so there are fewer free parameters and more counts per energy bin for the background model. 

\swift\ is slewing $\sim15-20\%$ of the time, so the inability to run NITRATES on data during these time periods has a significant impact on the recovery fraction. To date, GUANO data during slews has been analyzed using a bespoke mosaic imaging technique, similar in concept to that devised by \cite{BATSS}. However, as we have demonstrated imaging has crucial handicaps compared to NITRATES. The unstable attitude during slews however complicates the analysis substantially. One way to extend NITRATES to handle these periods would be by binning the data in time bins of $\lesssim0.2$ s, so that the pointing direction remains effectively unchanged during the bin. Then the likelihood could be evaluated over the 3D binning of detectors, energy, and time as apposed to the usual 2D. The signal model is often expressed as a function of detector coordinates ($\theta$, $\phi$) as sky coordinates are usually stationary compared to them, but now the model will need to be a function of sky coordinates (RA, Dec). When the LLH is maximized at a specific (RA, Dec), at each time bin there will need to be a different response for the new ($\theta$, $\phi$) that corresponds to that (RA, Dec) at that time. Between having to evaluate the LLH at an increased number of bins and needing to have several responses this version of the analysis would likely have a significantly higher computational cost. In addition, fitting the background model will also be more complicated as the background can quickly vary with pointing direction. The diffuse model would most likely need have time dependence added to it, and like the signal model, known bright sources that move across the FOV will need a new response for each time bin.  

Alternatively, the likelihood could be changed to an unbinned extended likelihood, which instead of summing the LLH over bins of data each event essentially has its own LLH that are summed and added with the Poisson probability of seeing the total number of counts. Each event's likelihood comes from ``background'' and ``signal'' probability densities as a function of its observables (energy, detector position, and time). This likelihood form has the benefit of not having the number of bins blow up, but would need work in implementing and testing it. It is also not clear how the model errors could be accounted for in this scenario. Despite these challenges, we plan to work on a slew-data capable version of NITRATES in the future, given the high yield of such a search.

\subsection{Computational Cost and Run Times}
As previously mentioned, this search has a relatively high computational cost. To complete the LLH optimization for a single time bin across the entire FOV (ignoring position seeding) takes $\approx$ 500 CPU hours. This increases approximately linearly with the number of time bins and decreases linearly with the fraction of the FOV that passes the position seeding. A typical search with an obvious signal may have two time bins and 20\% of the FOV that pass the seeding resulting in $\approx$ 200 CPU hours needed. The search is usually run on 100 cores, which means this would take about 2 hours to process (ignoring the time it takes to run the seeding and waiting in the cluster's queue). In practice the run time has significant variability since some bursts may have up to 6 time bins and $\approx$ 50\% of the sky that pass the seeding, resulting in run times closer to 8 hours. The search is embarrassingly parallelizable, and could be made significantly faster with access to any of: more CPU compute nodes, access to more storage to allow precomputing more forward models ($\sim10$ TB would be required) since the computational cost of constructing these forward models is on the same order of the likelihood computations, or optimization of the code for use on GPUs.

\section{Conclusions}
The NITRATES likelihood analysis and search we have described here effectively bypasses the limitations of conventional coded aperture imaging by utilizing all of the counts -- even uncoded -- on the detector array, the information associated with each count, and avoids the mask-weighting efficiency penalty. In addition, NITRATES allows the systematic discovery, characterization, and occasionally a rough localization, of OFOV GRBs with BAT for the first time, finally utilizing its high effective area even outside of the coded FOV. NITRATES has been implemented as a fully autonomous, low-latency targeted search, and in addition to GRB triggered searches has been run over a thousand times on GW, neutrino, and FRB triggers, setting the deepest possible GRB upper-limits for these events (which will be reported in future publications). NITRATES has demonstrated the ability to systematically detect and localize GRBs that are too weak for conventional imaging. It extends the detection and localization horizon for weak GRB 170817A-like bursts by greater than a factor of 1.5 compared to other searches and instruments, and thus dramatically increases the localization rate of nearby off-axis GRBs, with exciting implications for rapid follow-up of a gravitational wave source without lengthy wide-field optical searches. The increased arcminute-localized GRB yield from GUANO + NITRATES of $\sim13$ per year represents a substantial fraction of the total rate, and its higher short GRB recovery fraction as compared to BAT onboard will significantly extend our sample of localized cosmological short GRBs and their compact object progenitors.

\acknowledgements
We are very grateful to Takanori Sakamoto for providing access to the \swift\ Mass Model. We are delighted to thank J. Michael Burgess, Christopher Matzner, Brad Cenko, David Palmer, and Eric Burns for their helpful comments on an early version of this manuscript.
This material is based upon work supported by the National Science Foundation under Grants PHY-1708146 and PHY-1806854 and by the Institute for Gravitation and the Cosmos of the Pennsylvania State University. Any opinions, findings, and conclusions or recommendations expressed in this material are those of the author(s) and do not necessarily reflect the views of the National Science Foundation.

\software{All software used in this publication including the framework, search pipeline, and response functions are publicly available. The work was made possible by healpy \citep{healpy}, Matplotlib \citep{matplotlib}, NumPy \citep{numpy}, AstroPy \citep{astropy}, Numba \citep{Numba}, and Geant 4 \citep{Geant}.}

\appendix
\section{Building the Responses}
\label{DetResp}

\subsection{CZT Detectors}
\label{CZTdets}
When a photon interacts with the detector in a way that transfers at least some of its energy to an electron, that electron then collides with other atoms, subsequently freeing more electrons. This results in a cloud of electrons as well as a cloud of atoms missing electrons, known as holes. A voltage difference is applied between the top and bottom of the detector so that the electrons and holes migrate in opposite directions, creating a current. This current is measured and integrated over time to get the total charge, which is proportional to the energy deposited by the photon. This measured charge is converted to a pulse height amplitude (PHA) to be stored in the data. PHA values are converted to a measured photon energy using a polynomial with coefficients found for each detector using both pre-launch and onboard calibrations. 

In the energy range of interest ($\sim$ 10 keV - 1 MeV), the main ways photons interact with CZT detectors are through photoelectric absorption and Compton scattering. At energies $\lesssim$100 keV photoelectric absorption are the dominant type of interaction and at $\gtrsim$200 keV Compton interactions start to dominate. 

When a photon interacts through photoelectric absorption its energy is absorbed by the atom, which then ejects an electron. The electron is usually from the innermost shell, leaving the remaining electrons to rearrange to fill the vacancy, emitting a photon in the process. This new photon (florescence photon) will have a certain energy, pertaining to the transition the electron went through and will either interact with other atoms in the detector or escape the detector. This results in the original photon having either deposited its full energy into the detector or its full energy minus the energy of the escaped florescence photon. 

When a photon interacts through Compton scattering it collides into an electron imparting some of its energy into the kinetic energy of that electron. The photon then carries off with a new trajectory with a reduced energy and may or may not interact again inside the detector. If the photon interacts again it is possible that its full original energy ends up being deposited into the detector or just some fraction resulting in a continuum of possible deposited energies. 

The measurement of the charge in the detector is not a perfect practice, as it depends on the ability of the electrons and holes to successfully migrate through the detector. Defects in the detector cause charge trapping, where an electron or hole can become trapped for long enough that it is not included in the current readout resulting in a lower measured charge. Charge trapping affects holes more than electrons, and with voltage applied the holes travel to the top of the detector, so the deeper in the detector the interaction is the more charge is lost. 

The charge efficiency $\eta$ due to trapping follows the Hecht relation \cite{Hecht},

\begin{equation}
\eta(z) = \frac{Q}{Q_0} = \frac{\lambda_e}{D} (1 - exp \left(  - \frac{D-z}{\lambda_e}\right)) + \frac{\lambda_h}{D} ( 1 - exp \left( -\frac{z}{\lambda_h}\right) )
\label{Hecht}
\end{equation}
where $Q_0$ is the charge generated at depth z, $Q$ is the charge that makes it through the detector, $D$ is the thickness of the detector, and $\lambda_e$ and $\lambda_h$ are the average path lengths for electrons and holes. $\lambda_e$ and $\lambda_h$ are a function of the electric field applied $E$ and the detector's mobility-lifetime $\mu \tau_e$ and $\mu \tau_h$. 

\begin{equation}
\begin{split}
\lambda_e &= \mu \tau_e E \\
\lambda_h &= \mu \tau_h E \\
\end{split}
\end{equation}
$\mu \tau_e$ and $\mu \tau_h$ were found for each detector by the BAT team \citep{MuTfit}. In CALDB the mobility values were binned into 36 bins of $\mu \tau_e$ and $\mu \tau_h$ and only the bin centers and fraction of detectors per bin are provided to be able to calculate a detector summed response. 

Using the Hecht relation combined with knowledge of the probability distribution of photon interaction depths and energy deposition the DRM for a detector can be calculated. These distributions can be found through simulations of a photon flux on a detector. 

\subsection{Simulations}

SwiMM is used in two different ways here. In sections \ref{DetsOnly} and \ref{FullCraft} it is used to fully simulate a photon flux onto the model and in section \ref{TransSect} the positions, dimensions, and materials of the model parts are used to calculate the transmission alone (for the direct response). A photon flux of a specific energy  and from a given source position is simulated onto SwiMM, or part of SwiMM. Then, for each photon if any energy ends up being deposited into a detector; the total energy deposited, $E_{dep}$, the $E_{dep}$-weighted average detector depth, and which detector, in (detx, dety) coordinates is recorded. The $E_{dep}$-weighted average detector depth is recorded as opposed to the depth of each $E_{dep}$ to simplify the results due to storage limitations, and will be referred to as the depth through the rest of this section. 

\subsubsection{Detectors Only}
\label{DetsOnly}

For the purposes of determining the direct response it is sufficient to  use SwiMM with most of its components removed. Only the detectors and components close to the detectors, mainly the housing and electronics below each detector module, are retained. The effects of those nearby components and surrounding detectors are thus included, such as when a photon scatters off the housing and then into the detector, or when a photon does not reach a detector because it first hits and interacts in a neighbor detector. With this reduced mass model the response does not vary as quickly with respect to source position, and therefore allows a smaller simulation set. These simulations are run at several steps in detector spherical coordinates ($\theta$, $\phi$), with 23 steps in $\theta$ from 0$^\circ$ to 180$^\circ$ and 4 steps in $\phi$ from 0$^\circ$ to 45$^\circ$. The rest of $\phi$ is performed using simple rotations. 

The goal is to have a response for each detector. However, without the mask and other spacecraft and instrument components the response is very similar for many detectors. To decrease the number of simulations required and simplify the end product, the results for similar detectors are grouped together. The detector array consists of 256 sub-arrays of 8x16 detectors, with a 0.2 mm gap between detectors inside the sub-array. Consequently, the sides of the detectors along the edges of the sub-array are less blocked by neighboring detectors, resulting in a different response, especially for sources far from $\theta$ = 0. The results are therefore grouped into non-edge detectors, and the four sub-array edge detectors. At positions near perpendicular to the detector array, sub-arrays at the source-side of the detector array start to block the paths of photons to the other sub-arrays. This affects $\theta$'s $\approx$ 70$^\circ$ - 145$^\circ$. The asymmetry of this affect around the detector plane is due to the housing below the detectors. In addition to the previously mentioned group, the results are also grouped across similar sub-arrays due this effect. 

\begin{figure}[htb]
	\centering\includegraphics[width=5 in]{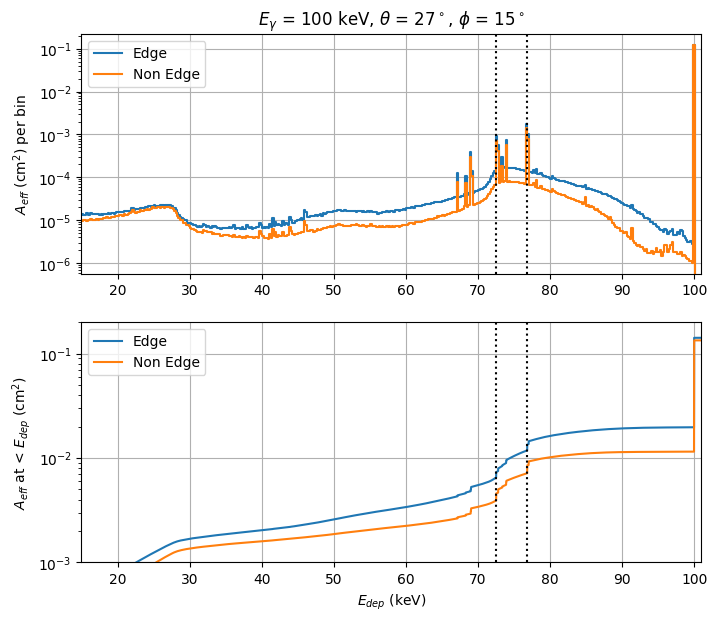}
	\caption{Both plots are for the detectors-only simulation with 100 keV photons and source position $\theta$ = 27$^\circ$, $\phi$ = 15$^\circ$. The top plot shows the single detector effective area divided over bins of $E_{dep}$ for both the edge and non-edge detectors. The 2 dashed lines show the main Cd and Te escape lines. The continuum can clearly be seen across all spectrum of $E_{dep}$, and the photo-peak line is clearly the largest line. The bottom plot shows the same value as the top plot but cumulative, better showing the relative contributions from the lines and continuum. }
	\label{E100Theta27Spec}
\end{figure}

\begin{figure}[htb]
	\centering\includegraphics[width=5 in]{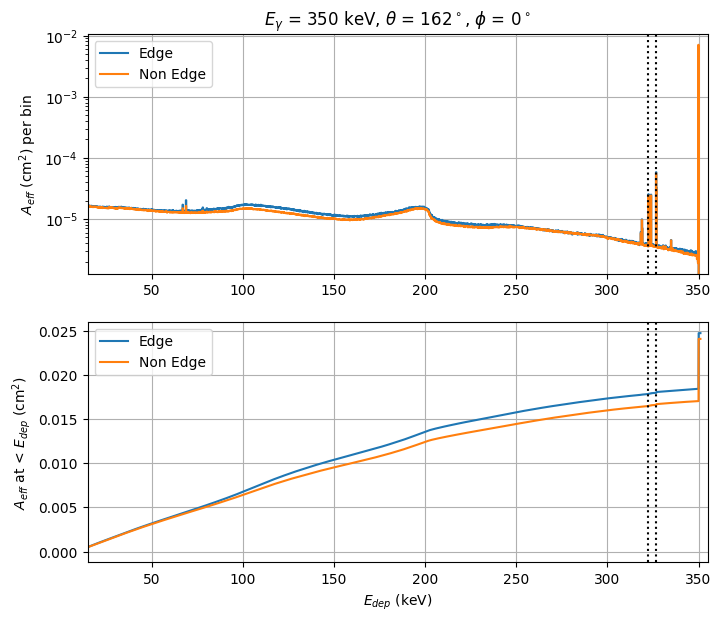}
	\caption{These are the same types of plots as in Figure \ref{E100Theta27Spec}, but for 350 keV photons and a source position of $\theta$ = 162$^\circ$, $\phi$ = 0$^\circ$. The continuum contribution to the total effective area is much higher here than for the previous case at $E_\gamma$ = 100 keV. }
	\label{E350Theta162Spec}
\end{figure}

\begin{figure}[htb]
	\centering\includegraphics[width=5 in]{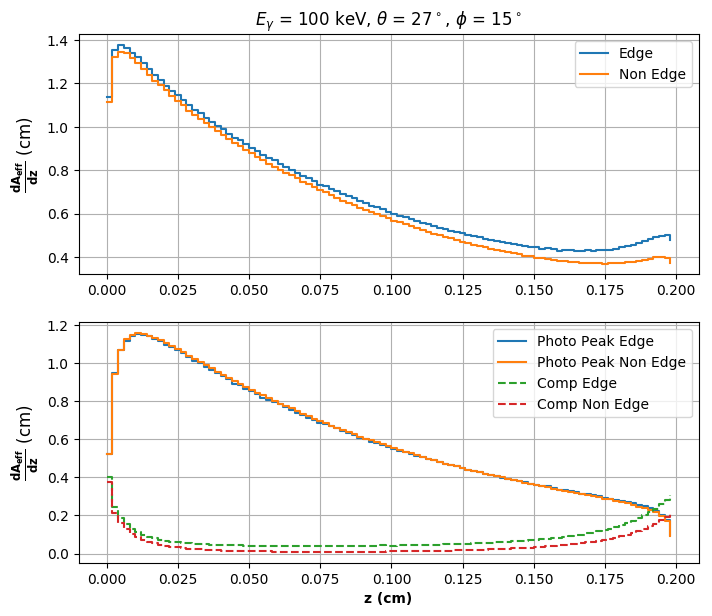}
	\caption{These plots are made from the same simulation as Figure \ref{E100Theta27Spec}, but what's plotted is $\frac{dA_{\rm eff}}{dz}$ as a function of depth. The top plot has $\frac{dA_{\rm eff}}{dz}$ for all $E_{dep}$'s plotted for both edge and non-edge detectors. The bottom plot has $\frac{dA_{\rm eff}}{dz}$ plotted for just the photo-peak line and for all $E_{dep}$'s in the continuum (marked "Comp" in the legend). The photo-peak distributions seem to be pretty much the same for edge and non-edge detectors, while for the continuum distributions there's a decent difference between edge and non-edge detectors. }
	\label{E100Theta27Depths}
\end{figure}

\begin{figure}[htb]
	\centering\includegraphics[width=5 in]{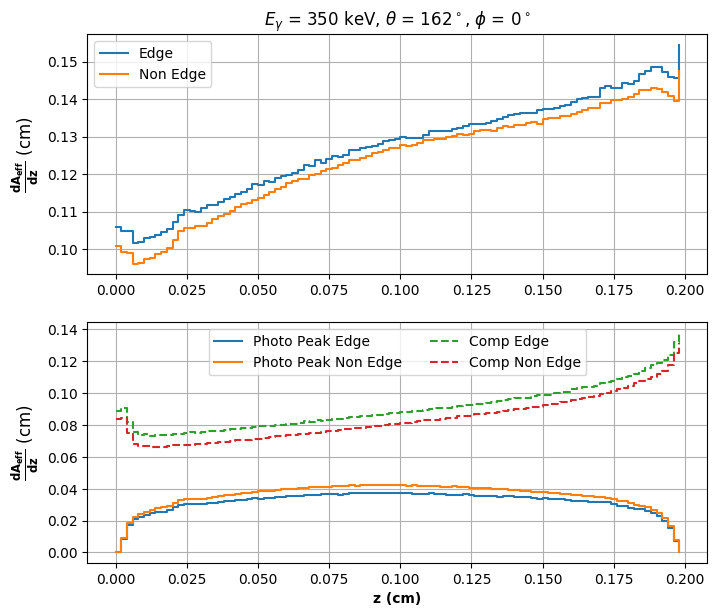}
	\caption{These are the same types of plots as in Figure \ref{E100Theta27Depths}, but with the same simulation used in Figure \ref{E350Theta162Spec}. The peak of the distribution is at the largest depth since the photons are now coming from beneath the detectors. In the bottom plot, differences between edge and non-edge can be seen for both the photo peak and continuum distributions. }
	\label{E350Theta162Depths}
\end{figure}

For this simulation there are 3 major $E_{dep}$ lines; the largest being the photo-peak line where the total photon energy, $E_{\gamma}$ is deposited into the detector and the most prominent escape lines from both Cd and Te. The escape energy for Cd and Te are 23.172 keV and 27.471 keV, so the $E_{dep}$ for the escape lines will be $E_{\gamma}$ - 23.172 keV and $E_{\gamma}$ - 27.471 keV. These lines can be seen in Figures \ref{E100Theta27Spec} and \ref{E350Theta162Spec}. For each group and each line the 1D depth distribution is calculated using the simulation products. This is done by selecting the events that have a $E_{dep}$ within 0.1 keV of the line energy and making a histogram binned in depth. The effective area would be the number of detected photons divided by the simulated photon fluence, so dividing the histogram bin counts by the photon fluence, the number of detectors in the group, and the bin width gives us the differential effective area, $\frac{dA_{\rm eff}}{dz}$ as a function of depth for that line. An example of this can be seen in the bottom of Figures \ref{E100Theta27Depths} and \ref{E350Theta162Depths}, where $\frac{dA_{\rm eff}}{dz}$ is plotted for the photo-peak line. 

A very similar process is done for the 2D depth distribution of the continuum, which is mostly from Compton scattering but includes everything else not in the lines. The continuum can be seen well in Figures \ref{E100Theta27Spec} and \ref{E350Theta162Spec}, where additional lines can be seen, but will be absorbed into the continuum here. For each group of detectors all the events with $E_{dep}$ that are not within 0.1 keV of one of the line energies are selected and binned in $E_{dep}$ and depth to make a 2D histogram. The histogram is divided by the photon fluence, number of detectors, depth bin width, and $E_{dep}$ bin width (correcting for the missing portions around the lines) to give us the differential effective area, $\frac{dA_{\rm eff}}{dzdE_{dep}}$ as a function of depth and $E_{dep}$. 

The depth distributions for lines and the continuum are made for each group of detectors and for each simulation done. A simulation for each source position is done at an array of $E_{\gamma}$'s ranging from 10 keV up to 6 MeV, with finer steps in $E_{\gamma}$ from 10 keV - 200 keV and steps getting larger the higher $E_{\gamma}$ goes.

\subsubsection{Full Craft}
\label{FullCraft}

\begin{figure}[htb]
	\centering\includegraphics[width=5 in]{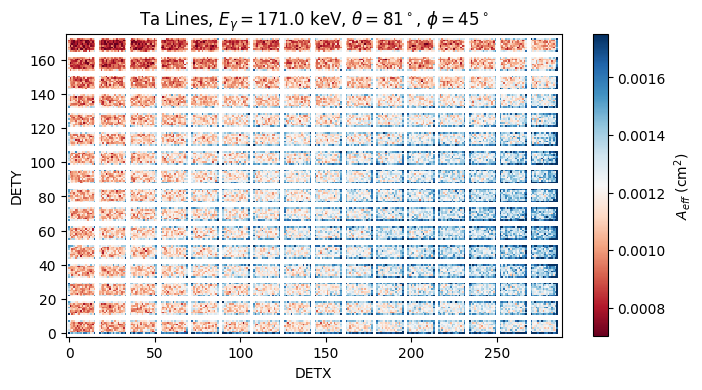}
	\caption{This figure uses simulated data from the full detector simulation with a photon energy of 171 keV and a source position of $\theta$ = 81$^\circ$, $\phi$ = 45$^\circ$. Plotted here is the effective area at the Ta florescence lines for each detector. This shows an example of the pattern that the florescence lines create across the detector array.}
	\label{E171Theta81TaDPI}
\end{figure}

To calculate the indirect response simulations are performed using the full SwiMM. Just like in section \ref{DetsOnly} there will be lines and a continuum, but this time the lines originate from fluorescence in the shield and other material surrounding the detectors. With the full model the simulation takes longer to run, so the edges are not grouped separately. Each detector sub-array is treated as its own group as the pattern of photons for both the lines and continuum changes across the detector array, as can be seen in Figure \ref{E171Theta81TaDPI} with the florescence lines of Ta. Unfortunately there is no symmetry to take advantage of here, so the incident angles for the simulated photon flux must cover the whole sky. Sky positions are chosen based on a HEALPix map with NSide=4, and additionally at $\theta$'s 0$^\circ$ and 180$^\circ$. 

The graded-Z fringe shield surrounds the instrument, closing out the 1 meter separation between the detector plane and the coded mask, and also shields the underside of the detector plane. The shield is comprised of layers of Pb, Ta, Sn, and Cu with varying thickness between the detector array and the mask. Cu has no florescence lines above 10 keV; this is below the low-energy threshold of the detectors, so they are safely ignored. For the other layers the most prominent lines are used; 73.03 keV, 75.25 keV, 84.75 keV and 85.23 for Pb, 56.41 keV, 57.69 keV, 65.11 keV, 65.39 keV, 67.17 keV for Ta, and 25.03 keV, 25.25 keV, 28.47 keV for Sn. The depth distributions are made using the same method from section \ref{DetsOnly}. 

The 2D depth distributions are made also made using the same method from section \ref{DetsOnly}, except both the lines from section \ref{DetsOnly} and the florescence lines are removed. 


\begin{figure}[htb]
	\centering\includegraphics[width=5 in]{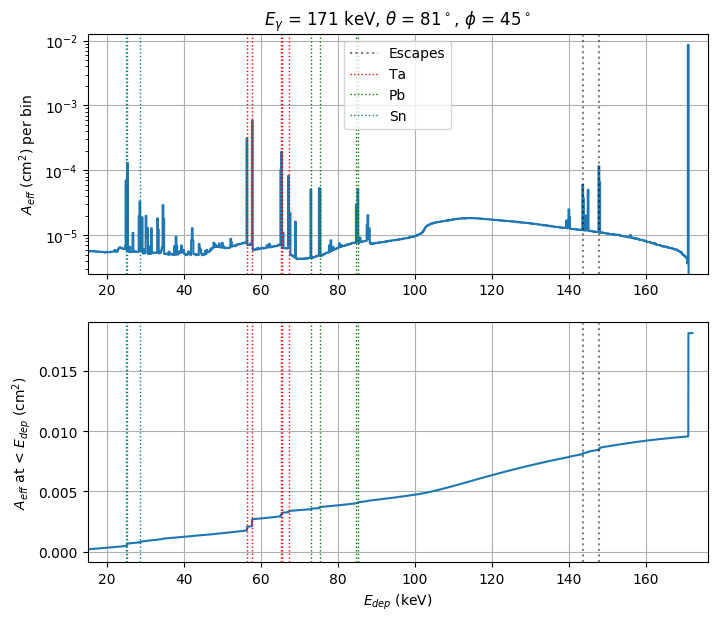}
	\caption{These plots show the same plotted values as Figures \ref{E100Theta27Spec} and \ref{E350Theta162Spec}, but for a full detector simulation with a photon energy of 171 keV and source position of $\theta$ = 81$^\circ$, $\phi$ = 45$^\circ$. In addition to the vertical lines plotted for the escape lines, there are also vertical lines plotted for the florescence lines.}
	\label{E171Theta81Spec}
\end{figure}

Figure \ref{E171Theta81Spec} shows the distribution of $E_{dep}$'s for a simulation with $E_\gamma$ = 171 keV, $\theta$ = 81$^\circ$, and $\phi$ = 45$^\circ$. The top plot shows the size of all of the included florescence lines, which are large but not quite as large as the main photo-peak. It also shows the size and shape of the continuum, which peaks around an $E_{dep}$ of 110 keV and flattens out near 60 keV. The bottom plot shows the cumulative effective area with $E_{dep}$, where the small jumps at each line can be seen along with the $\approx$ half of the $A_{dep}$ coming from the continuum and florescence lines with the other half coming from the photo-peak. 

One issue with this simulation for the indirect continuum is that it contains the direct continuum response within it. After the simulations were run it was not possible to remove events that first interacted in or near the detector. To avoid double counting this in the full response only one will be able to be used. For source positions inside the coded FOV the Compton scattering off of the craft is a minor part of the response, so we choose to ignore it. For positions outside the coded FOV the response won't change significantly at an arcminute scale like it does when modulated by the mask, so using the indirect continuum response will work well. The divide between which continuum response is used is placed at $\theta$ = 70$^\circ$, which is the lowest $theta$ where a simulation was ran with the full craft before some $\phi$'s were inside the FOV. For some $\phi$ angles the coded FOV does not begin until $\approx$ 45$^\circ$, so unfortunately at these positions the continuum response may be pretty poor. This affects $phi$ around 270$^\circ$ the most as this is where the photons will have to go through the XRT and UVOT tubes causing a large amount of Compton scattering compared to the number of photons traveling through them unimpeded. This deficiency could possibly be remedied with more simulation in the future.

\subsection{Precomputed DRMs}
\label{PreComputeSect}

Now with the depth distributions and $\mu \tau$ values on hand the precalculated part of the DRMs can finally be calculated. For the line responses the DRM is calculated in a very similar way to what is done in the ftool \textit{batdrmgen}. For a single 1D depth distribution we want to find how effective area is split between the energy bins, $R_j$ from equation \ref{DRMrow}. We do this by going though each depth bin to find the effective area and the `effective energy', $E_{\rm eff}$ in that depth bin and adding the effective area to the energy bin for $E_{\rm eff}$. The depth bin's effective area and $E_{\rm eff}$ is given by,

\begin{equation}
\begin{split}
A_{\rm eff} (z, \Delta z) &= \Delta z \frac{dA_{\rm eff}}{dz}(z) \\
E_{\rm eff}(z) &= A_{gain} E_{dep} \frac{\eta(z)}{max[\eta]}
\end{split}
\end{equation}
where $E_{dep}$ is the line energy, $\Delta$z is the height of the depth bin, $A_{gain}$ is a near unity adjustment factor from calibration, and $max[\eta]$ is the maximum possible value of $\eta$ for the $\mu \tau$ values being used. The summed effective area for each energy bin is then convolved with a Gaussian to take into account the measurement resolution, where the energy-dependent $\sigma$ value was found during calibration. This gives us $R_j$ for a specific set of $\mu \tau_e$ and $\mu \tau_h$, so this is done for each set given in the CALDB table and the weighted average is taken using the listed fractions as the weights. 

For the continuum responses the process is very similar. For a single 2D depth distribution an array of $E_{dep}$ values are chosen going from 10 keV to just below $E_\gamma$, with spacing $\Delta E_{dep}$. At each $E_{dep}$ value the 1D depth distribution is found by doing a linear interpolation of the $\frac{dA_{\rm eff}}{dzdE_{dep}}$ values at the adjacent $E_{dep}$ bins and multiplying it by $\Delta E_{dep}$. The same process to find $R_j$ for the line responses is then used to find $R_j$ for each of these $E_{dep}$ values. Then, the $R_j$'s are summed together. 

For each position the simulation was run at, these $R_j$'s are made for each $E_\gamma$ to make the DRM matrix. These DRMs are stored and used in the analysis by being interpolated between spatially to get the response at a specific position. The direct response for each detector still needs the transmission probability taken into account, but this doesn't effect the spread across energy bins for a specific $E_\gamma$, it just effects the effective area.


\subsection{Transmission}
\label{TransSect}

For the direct DRMs, the transmission probability from the source to the detector has been ignored. Here it will be shown how it is calculated and also how it is calculated for the special case of through the coded mask. This information is added into the precalculated direct DRMs like this,

\begin{equation}
DRM^{direct}_{ijl} = t_{il} DRM^{precalc,direct}_{ijl}
\end{equation}
where $t_{il}$ is the transmission probability for a photon with the $l$'th energy to the $i$'th detector and ${\rm DRM}^{precalc,direct}_{ijl}$ is the precalculated portion of the direct DRM. 

\subsubsection{Through the Craft}

The positions, dimensions, and materials for the major pieces of the spacecraft are taken from SwiMM and stored in custom Python objects. The goal is to be able to quickly calculate the transmission probability, $t_i$ for a photon from the source position to a detector. To do this we need the distance, $d$ traveled through each part of the craft, the density, $\rho$ of each of those parts, and the mass absorption coefficient $\mu$ for each of those parts, which is a function of $E_\gamma$. $t_i$ through one of those parts can then be calculated as 

\begin{equation}
t_i = e^{-\mu \rho d}
\end{equation}
and though all of the parts $t_i$ is the product of the $t_i$ for each part. 

The mass absorption coefficient curves as a function of photon energy for each of the element components in the materials found in SwiMM are taken from the `XCOM: Photon Cross Sections Database' \cite{XCOM} on the NIST website. 

The distances through each part are found by first finding which parts the photon path will intersect, if any. Then, the entrance and exit positions are found using ray-tracing techniques and the distance is just the distance between those two positions. This is done for each detector quickly using a Python function optimized using Numba. 

\begin{figure}[htb]
	\centering\includegraphics[width=6 in]{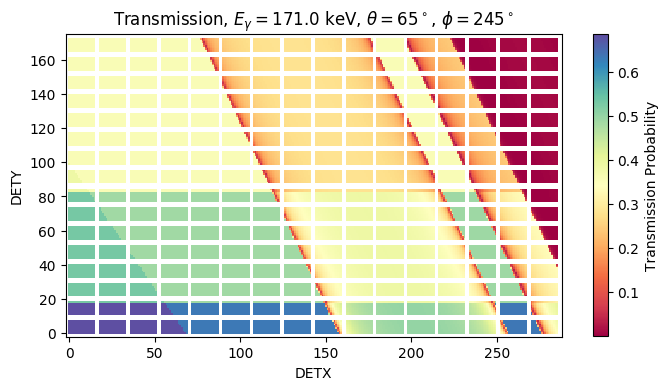}
	\caption{The calculated transmission probabilities from the source to each detector for a photon with energy 171 keV, coming from a source at $\theta=65^\circ , \phi=245^\circ.$}
	\label{TransDPI}
\end{figure}

\subsubsection{Through the Mask}


When the path of a photon goes through the mask the many lead tiles need to be taken into account, which would take way too long using the method for the other parts of the craft due to the high geometric complexity. Instead, an ftool called \textit{batmaskwtimg} is used to calculate the fraction of each detector that would be not shadowded by the mask tiles, $f_i$ using a built-in ray-tracing algorithm. There are a lot of positions that need to be done with the large FOV and much smaller PSF, and so all of them are precomputed. In tangential plane coordinates (imx, imy), the PSF FWHM is $\approx$ 0.0065, so the ray-tracings will be done on an imx, imy grid with spacing of 0.002, which is similar to the pixel widths of a BAT image made using default settings. To be able to get $f_i$ for any position, a bi-linear interpolation in imx, imy coordinates with the 4 nearest ray-tracings is used. 

\begin{figure}[htb]
	\centering\includegraphics[width=6 in]{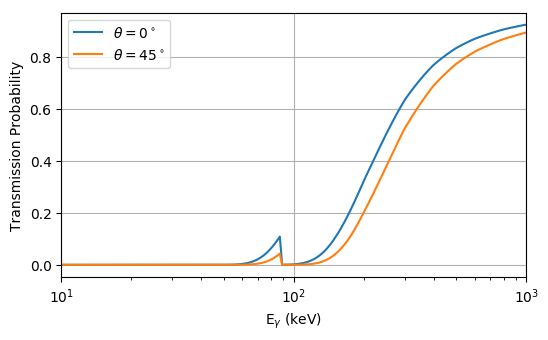}
	\caption{The transmission probability of a photon through a lead tile the BAT coded mask as a function of photon energy. }
	\label{PbTrans}
\end{figure}

Even though the purpose of the lead tiles is to block photons from getting through, they still have a finite probability of transmission, especially at higher energies as shown in Figure \ref{PbTrans}. For the parts of a detector that is shadowed by a lead tile the path could come out the side of the tile or enter through the side, but to simplify things that is ignored. So, the distance through that tile is just assumed to be the tile height (1 mm) divided by $cos(\theta)$. $t_i$ can then be calculated as,
\begin{equation}
t_i = (t_i)_{\rm non-mask}\times(f_i + (1 - f_i)t_{pb})
\label{Tmask}
\end{equation}
where $(t_i)_{non-mask}$ is the transmission probability through anything besides the lead tiles (like the mask support structure) and $t_{pb}$ is the transmission probability through the lead tile.

\subsection{Putting It All Together}
\label{PuttingSect}

With the precomputed portions of the response derived, the full response for the analysis can now be assembled. Since these were made at a grid of positions, when computing the response for a specific position point a linear interpolation of the precomputed DRMs will be used. The transmission probabilities will be calculated for the exact position. The full DRM for the $i$th detector can then be calculated as,

\begin{equation}
\begin{split}
\text{when $\theta \leq$ 70$^\circ$: } \\
DRM_{ijl} &= t_{il} (DRM^{\rm lines,direct}_{ijl} + DRM^{\rm cont,direct}_{ijl}) + DRM^{\rm lines,indirect}_{ijl} \\
\text{when $\theta >$ 70$^\circ$: } \\
DRM_{ijl} &= t_{il} (DRM^{\rm lines,direct}_{ijl}) + DRM^{\rm cont,indirect}_{ijl} + DRM^{\rm lines,indirect}_{ijl} \\
\end{split}
\end{equation}
where ${\rm DRM}^{x,y}_{ijl}$ are the interpolated precomputed DRMs.

The appropriate response error is not clear from simulation alone. For a mask-weighted spectra the error derived by the BAT team is 4\%, which is found after thorough calibration on observational data. The error here should be at least that much and the more complicated the photon path is the more chances there are for error. We set the error for direct lines at 10\% and the error for both florescence lines and the continuum at 16\%. These values are not well founded, but they have the desired effects of; not being too small, not being so large that it messes up the likelihood (probability of a negative effective area should be very small), and the direct lines response has a smaller error. Through future calibration on bright bursts a more robust error model can be made.

\begin{figure}[htb]
	\centering\includegraphics[width=6 in]{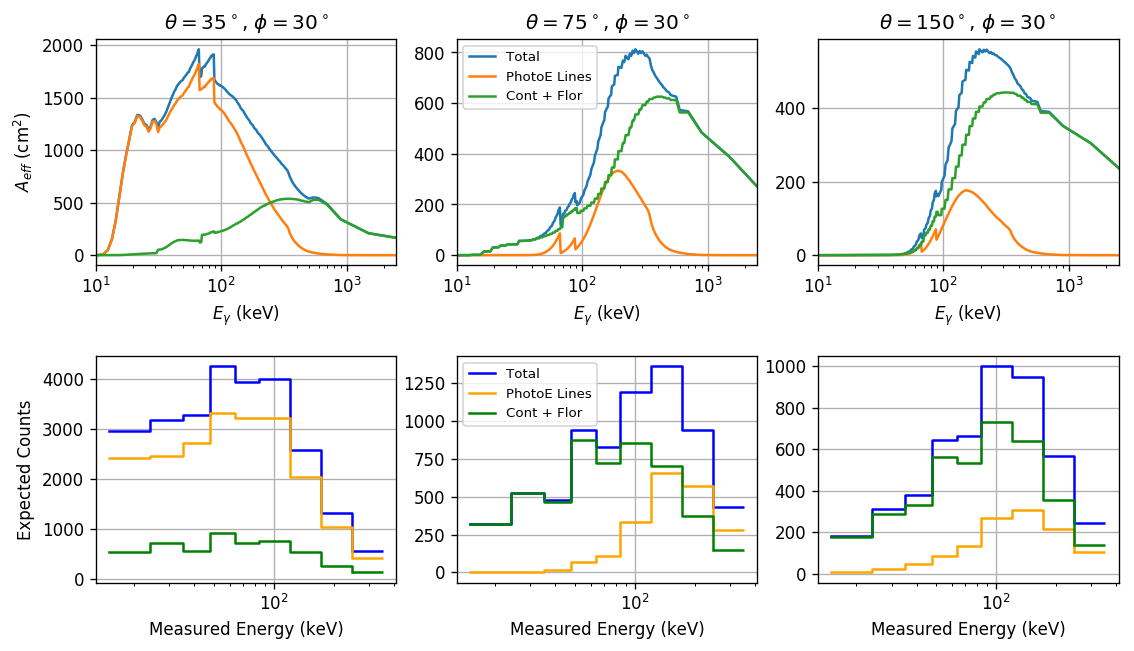}
	\caption{The top row of plots here show the total effective area over all 32,768 detectors calculated by the response as a function of $E_\gamma$ at three different $\theta$ angles. The plots show the total effective area as well as split into the photo-peak plus escapes lines and continuum plus florescence lines. The bottom row of plots shows the expected number of counts over all detectors split into the 9 energy bins used in the analysis for a high-energy exponentially cutoff powerlaw spectra with $\gamma$ = 0.5, $E_{\rm peak}$ = 350 keV, and a 10 keV - 1000 keV fluence of 5 $\times$ 10$^{-6}$ erg cm$^{-2}$. The bottom plots are also split into counts from photo-peak plus escapes lines and continuum plus florescence lines and are for the same source positions as the plots above them.}
	\label{SpecAeff}
\end{figure}

With the total response calculated now, we can explore the response across positions and energies. In the top row of Figure \ref{SpecAeff}, the total effective area for all detectors is plotted as a function of $E_\gamma$ for three different positions; $\theta$ = 35$^\circ$, which is in the coded FOV, $\theta$ = 75$^\circ$, which is outside the coded FOV, and $\theta$ = 150$^\circ$, which is also outside the coded FOV but coming from below the detector array. It can be seen that the effective area drops significantly from $\theta$ = 35$^\circ$ to $\theta$ = 75$^\circ$ and drops some more going to $\theta$ = 150$^\circ$. It is also shown how effective area is split between the direct lines and the florescence plus continuum. Within the FOV the effective area is dominated by the photo-peak and escape lines and peaks at below 100 keV, while outside the FOV the effective area is dominated by the continuum and florescence lines and peaks at $\approx$ 200 keV. The bottom row of Figure \ref{SpecAeff} shows the number of expected counts split into 9 energy bins for a fluence of 5$\times$10$^{-6}$ erg/cm$^2$ following an exponentially high-energy cutoff powerlaw spectra with $\gamma$ = 0.5 and $E_{\rm peak}$ = 350 keV. As expected from the effective area curves, the number of expected counts is dominated by the response to photo-peak and escape lines inside the FOV and dominated by the response to the continuum and florescence lines outside the FOV. This spectra is on the harder side of normal GRBs, but there are still a significant amount of expected counts at lower energy bins for $\theta$ = 35$^\circ$. At the two higher $\theta$'s there are significantly less expected counts at the lower energy bins, but not a negligible amount with almost all of it coming from the continuum and florescence response. 
 

\begin{figure}[htb]
	\centering\includegraphics[width=5.5 in]{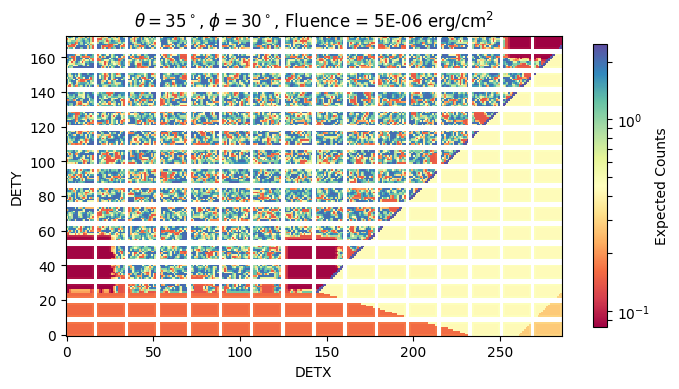}
	\caption{For the same spectra and fluence used in the bottom row of Figure \ref{SpecAeff}, this plot shows the distribution of expected counts across the detectors for a source position of $\theta$ = 35$^\circ$, $\phi$ = 30$^\circ$. The majority of the detector array is coded (photon path through the mask), and has large fluctuations in neighboring detectors due to the coded mask. The uncoded regions have large patches of the same expected counts, with changes when the photon path goes through a different part of the shield. }
	\label{CountsDPItheta35}
\end{figure}

\begin{figure}[htb]
	\centering\includegraphics[width=5.5 in]{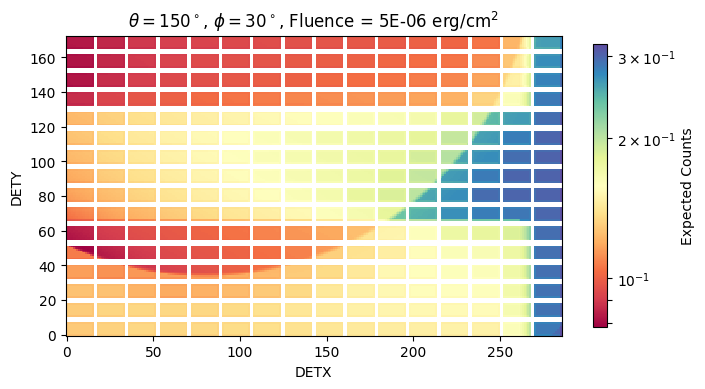}
	\caption{For the same spectra and fluence used in the bottom row of Figure \ref{SpecAeff}, this plot shows the distribution of expected counts across the detectors for a source position of $\theta$ = 150$^\circ$, $\phi$ = 30$^\circ$. The whole detector array is uncoded with the majority of the spatial pattern coming from the parts of the space craft the photon path goes through. }
	\label{CountsDPItheta150}
\end{figure}

Figures \ref{CountsDPItheta35} and \ref{CountsDPItheta150} show the distribution of expected counts in each detector for the same flux used in Figure \ref{SpecAeff}. In Figure \ref{CountsDPItheta35} the difference between the coded (photon path went through the mask) portion of the detector array and the uncoded portion is very apparent, with the counts in the coded portion having large fluctuations and the counts in the uncoded region having large areas of similar counts pertaining to which portion of the shield the photons traveled through. There's a line of large counts that run along the diagonal border of the coded and uncoded regions that comes from a small gap between the mask and shield. In Figure \ref{CountsDPItheta150} the expected number of counts have a much smaller range, with the spatial pattern coming mostly from the different parts of the craft the photons had to travel through.

\begin{figure}[htb]
	\centering\includegraphics[width=5 in]{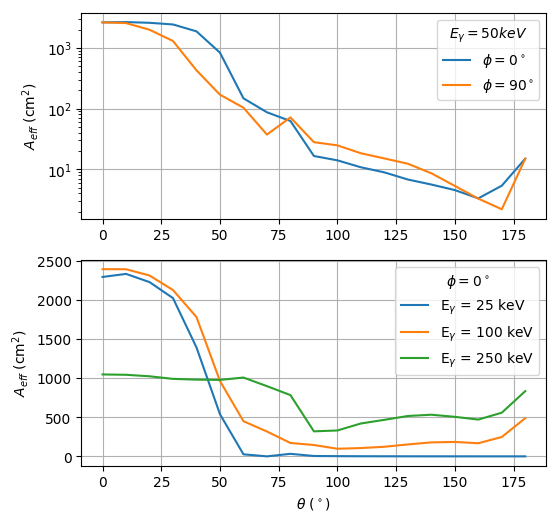}
	\caption{The top plot shows the calculated response's effective area for a photon energy of 50 keV as a function of $\theta$ for two different $\phi$ angles. The bottom plot shows the calculated response's effective area for three different photon energies all at a $\phi$ = 0$^\circ$. }
	\label{AeffVsTheta}
\end{figure}

The top of Figure \ref{AeffVsTheta} shows effective area at $E_\gamma$ = 50 keV as a function of $\theta$ for two different $\phi$'s. The large difference at $\theta$'s 20$^\circ$ to 60$^\circ$ comes from the partial coding fraction changing faster at $\phi$ = 90$^\circ$ than it does at 0$^\circ$. At $\theta$'s larger than 60$^\circ$, effective area at both $\phi$'s drop off to very low values. In the bottom of Figure \ref{AeffVsTheta} effective area as a function of $\theta$ is shown for three different $E_\gamma$ values. It can be seen that effective area at the two lower energies are much higher at $\theta$'s that are near the center of the FOV, then drops off quickly with larger $\theta$ where effective area at $E_\gamma$ = 250 keV becomes larger. 

\bibliography{refs}{}
\bibliographystyle{aasjournal}



\end{document}